%% file: main.tex
\documentclass[aps,preprint,amsmath,amssymb,groupedaddress]{revtex4-2}
\usepackage{CJK}
\usepackage{graphicx}

\usepackage{amsmath}
\usepackage{mathtools}
\usepackage{amsfonts}
\usepackage{braket}
\usepackage{comment}
\usepackage{subfigure}

\usepackage{tikz}

\begin{document}
\title{Failure detection for transport networks}
\author{Edoardo Rolando}
\affiliation{Department of Physics Freie Universit\"at Berlin\\}
\author{Armando Bazzani}
\thanks{Corresponding author: armando.bazzani@unibo.it}
\affiliation{Department of Physics and Astronomy Bologna University}
\affiliation{INFN sezione di Bologna\\}
\date{\today} 

\begin{abstract}
Diffusion on complex networks is a convenient framework to simulate a great variety of transport systems. The effects of failures in the network links may be used to cascade phenomena or the congestion formation in the system. A real time detection of failures can mitigate their effect and allow to optimize the control procedures on the transport network. The main objective of this work is to provide a dimensionality reduction technique for a transport network where a diffusive dynamics takes place, to detect presence of a failure by a limited number of observations. Our approach is based on the susceptibility response of the network state under random perturbations of the link weights. The correlations among the nodes fluctuations is exploited in order to provide the clustering procedure. The network dimensionality is therefore reduced introducing `representative nodes' for each cluster and generating a reduced network model, whose dynamical state is detected by the limited observations. We realize a failure identification procedure for the whole network, studying the dynamics of the coarse-grained network. The localization efficiency of the proposed clustering algorithm, averaging over all possible single-edge failures, is compared with traditional structure-based clustering using different graph configurations. We show that the proposed clustering algorithm is more sensitive than traditional clustering techniques to detect link failure with high stationary fluxes. 
\end{abstract}

\keywords{Diffusion on weighted graph, Clustering methods, Transport networks,  coarse-grained models}

\maketitle

\input{sections/section01.tex}

\input{sections/section02.tex}

\input{sections/section03.tex}

\input{sections/section04.tex}

\input{sections/section05.tex}

\input{sections/section06.tex}
\input{sections/section07.tex}
\input{sections/acknowledgements.tex}
\clearpage

\bibliography{bibliography}  

\appendix\input{sections/appendix.tex}

\end{document}

%% file: sections/section01.tex
\section{Introduction} \label{sec:introduction}
Network science has proven to be one of the successful tools to study the properties and the evolution of complex systems\cite{newman_networks_2018}. The diffusion processes are among the simplest dynamical systems on a graph structure that may describe a variety of phenomena in chemistry, biology and social systems providing a convenient framework to model different architectures or structures. \cite{spatial_networks}\cite{dynamical_processes_on_complex_networks}. 
For instance, in social sciences the diffusive dynamics may simulate the spreading of a disease or an opinion \cite{harary_social}, or in biological systems the so called Elastic Networks Models describe the proteins or gene interactions networks\cite{edge-based-elastic}\cite{prediction-of-allosteric-sites}\cite{mason_graph_2007}. Diffusion processes on graphs has also found fruitful applications for urban mobility models, water distribution systems\cite{wdn_spectral}, heat transport and power grid networks \cite{nonlocal_impact_of_link_failure}. One of the main issue remains a deep understanding of the interplay between the network structure and the diffusion dynamics. 
Focusing on transport systems, a typical problem is the study of their resilience and stability under the failure of one or more components. Even a single-edge failure can lead to a cascade effect (e.g congestion propagation), reducing the transport capabilities of the whole structure \cite{cascade_failure_Mirzasoleiman,cascade_failure_Chen,gonzalez2023}. In the literature, several studies focus on the definition and planning of optimal transport networks, such that the resilience and stability under failures is maximized \cite{barabasi2016,moutsinas2020}. But when the network structure is fixed, one has to cope with the problem of real time monitoring a transport system in order to detect the appearance of a failure and to apply real-time strategies that may mitigate the failure consequence taking into account the dynamical properties. This problem can be tackled by building an efficient sensor system, that guarantees the observability and localization of the failure with a limited spatial error. However,
monitoring all the components of transport systems may be unfeasible or expensive, so that one has to develop efficient procedures to detect the failure location using a partial knowledge of the dynamical state. At this purpose, traditional clustering algorithms (like spectral clustering, Girvan-Newman algorithm and modularity maximization) might be exploited in order to group the nodes into homogeneous communities\cite{fortunato_community_2010}.  All these methods rely on the definition of a cluster as a subset of nodes strongly connected and they are based on the topological properties of the network. Moreover, there is no natural choice how to select nodes to monitor each cluster, even if heuristic arguments might be made, exploiting traditional network centrality measures \cite{wdn_spectral}. However when the network structure supports a dynamics, like a diffusion process, the clustering procedure has to consider the dynamical properties of the system\cite{kurata2010}.\par\noindent 
In this paper we introduce a algorithm for the reduction of the network dimensionality, based on the susceptibility response of the nodes under a perturbation, when a diffusion dynamics takes place driven by an external forcing. Firstly, we add a white noise to the external forcing of the system and we show that the correlation properties depend on the spectral properties of the Laplacian matrix associated the diffusion dynamics. Then we study the effect of a perturbation of the link weights and we prove that correlation properties depend both on the spectrum of the Laplacian matrix and on the fluxes of the transport system. In the last case the eigenvectors able to distinguish edges with higher flux play a more relevant role in the clustering procedure. We exploit the correlation matrix to define a clustering procedure and we compare the efficiency under a single-edge failure event of the proposed clustering algorithm, with the traditional structure-based ones. The comparison is performed by means of a failure identification function using the dynamics occurring on a coarse-grained reduction of the initial graph. The network structures considered to test our clustering procedure are the grid network, the Erods-Renyi random network, the regular network and the Barabasi-Albert free scale network (the results for last two cases are reported in the Supplementary Material). It is shown that the proposed algorithm is more sensitive to failure of edges with higher stationary flows, that are the most important for a transport system. \par\noindent
A relevant application of our approach can be the traffic dynamics on an urban road network when local congestion arises and may spread on the road network\cite{gonzalez2023}. The vehicle interaction reduces the travel velocity on a road, and consequently the flow, as a function of the density (flow-density fundamental diagram), Therefore, when the density overcomes a critical value the equilibrium solution with uniform vehicle density becomes unstable and there is a sudden drop down of the traffic flow. In a typical situation, there is the possibility of measuring the dynamical state of a limited number of roads using distributed traffic flow sensors, like magnetic coils and video-cameras, and the development of real-time strategies to mitigate the congestion effects is one of the goals of the development of digital twins for urban mobility\cite{yeon2023}.\par\noindent
The paper is organized as follows: in the second and third sections we briefly introduce the diffusion processes on networks and discuss the effect of a single link failure. In the fourth section we study the system susceptibility under external perturbations and we compute the correlation matrices among the node states. In the fifth and sixth sections we introduce the proposed failure identification procedure and in the seventh section numerical results are shown to compare its efficiency with analogous procedures based on different clustering methods. Finally some conclusions are drawn. In the appendices we report some detailed calculations of the results discussed in the paper.

%% file: sections/section02.tex
\section{Diffusion processes on network and transport systems}\label{sec:diffusion}
We consider a weighted graph denoting by $w_{ij}\ge 0$ the weight of the link $j\to i$: we recall that if $w_{ij}$ is symmetric the graph is bidirectional. The diffusion process is defined by the graph Laplacian matrix 
\begin{equation}\label{def:laplacian}
	L_{ij} = 
	\begin{cases} 
		-w_{ij} & \text{ if } i \neq j \text{ and } j \text{ is connected to } i, \\
		d_{i}=\sum_j w_{ji}\quad & \text{ if } i=j, \\
		0 & \text{ otherwise }
	\end{cases}
\end{equation}
where $d_i$ is the weighted degree of node $i$. In matrix notation $L=D-A$, where $D$ is the diagonal degree matrix. 
Let $p_i\ge 0$ the state of the $i$-node (i.e. the particle density at the node $i$ of the diffusion process at the time $t$), $w_{ij}p_j$ defines the average flow through the link $j\to i$ and we have the master equation   
\begin{equation}\label{eq:our model}
\begin{aligned}
	\dot p_i =& \sum_j (w_{ij}p_j - w_{ji}p_i)+s_i \\
                 =& - \sum_j L_{ij} p_j + s_i
\end{aligned}
\end{equation}
where $s_i$ defines the node feature as a sink ($s_i<0$) or a source ($s_i>0$) and the sink-source features define the transport demand in the network realized by the diffusion process.I
n the sequel we assume $L_{ij}$ symmetric and eq. (\ref{eq:our model}) can be written in the form
$$
\dot p_i = \sum_j w_{ij}(p_j - p_i)+s_i \\
$$
so that the total flow on a link is proportional to the density difference (\textit{Fick’s law}). 
Throughout the text we will call the $s_i$ the \textit{demands} or \textit{loads} of the transport network. We remark that the diffusion dynamics (\ref{eq:our model}) is linear since the weights are constant, but nonlinear models can be considered by introducing dependence of the weights on the nodes states: e.g. $w_{ij}=w_{ij}(p_j)$ may describe the
congestion effect on the link $j\to i$ that reduces the transport capacity when the density $p_j$ overcomes a threshold, according to the existence of a fundamental diagram for the traffic dynamics\cite{daganzo2008,gonzalez2023}. In such a case the linear model (\ref{eq:our model}) can approximate the dynamics near a stationary solution and the failure event corresponds to a sudden drop in transport capacity of a link. 
The existence of a stationary solution for the system (\ref{eq:our model}) requires a balance between sources and sinks (assuming $s_i$ constant), so that the condition holds
\begin{equation}\label{eq:solvability condition}
	\sum_i s_i=0.
\end{equation}
For a connected graph the Laplacian character of the matrix $L_{ij}$ (i.e. $\sum_i L_{ij}=0$ and $L_{ij}\le 0$ if $i\ne j$) implies that the hyperplane $\sum_i p_i=0$ is an invariant space, and all the eigenvectors except the null one, belong to this hyperplane and have positive eigenvalues. The condition (\ref{eq:solvability condition}) implies that  $\vec s$ belongs to the invariant hyperplane where $L$ is invertible. The generic solution of eq. (\ref{eq:our model}) reads
\begin{equation}\label{eq:transient solution}
	\vec{p}(t) = \sum_{\lambda \neq 0} c_{\lambda} e^{-\lambda t} \vec{v}_{\lambda} +  {\vec p}^{\phantom{p}s}
\end{equation}
where ${\vec p}^{\phantom{p}s}$ is the average stationary solution and the coefficients $c_{\lambda}$ are determined by the initial conditions
\begin{equation}
	\vec{p}(0) = \sum_{\lambda \neq 0} c_{\lambda} \vec{v}_{\lambda} + {\vec p}^{\phantom{p}s}.
\end{equation}
The reciprocals of the eigenvalues $\lambda^{-1}$ define the relaxation time scales of the system towards the stationary solution. 
For $t\to\infty$ the first term in eq. (\ref{eq:transient solution}) decays to zero and the system relaxes to the stationary solution ${\vec p}^{\phantom{p}s}$ which satisfies
\begin{equation}\label{eq:stationary}
	L {\vec p}^{\phantom{p}s} = \vec{s}
\end{equation}
Explicitly we have 
\begin{equation}\label{eq:stationary solution}
	p_i^{s}=\sum_{\lambda\ne 0} \frac{s_i^\lambda}{\lambda} + p_i^0
\end{equation}
where $s_i^\lambda$ are the components of $\vec s$ on the eigenvector $\vec{v}^\lambda$. However the solution is not unique since one can add an arbitrary kernel component to ${\vec p}^{\phantom{p}s}$. To get an unique solution we require $\sum_i  p_i^s=0$ so that the stationary solution lies in the same subspace of the external forcing $\vec s$.

%% file: sections/section03.tex
\begin{figure*}[t]
    \includegraphics[width=\textwidth]{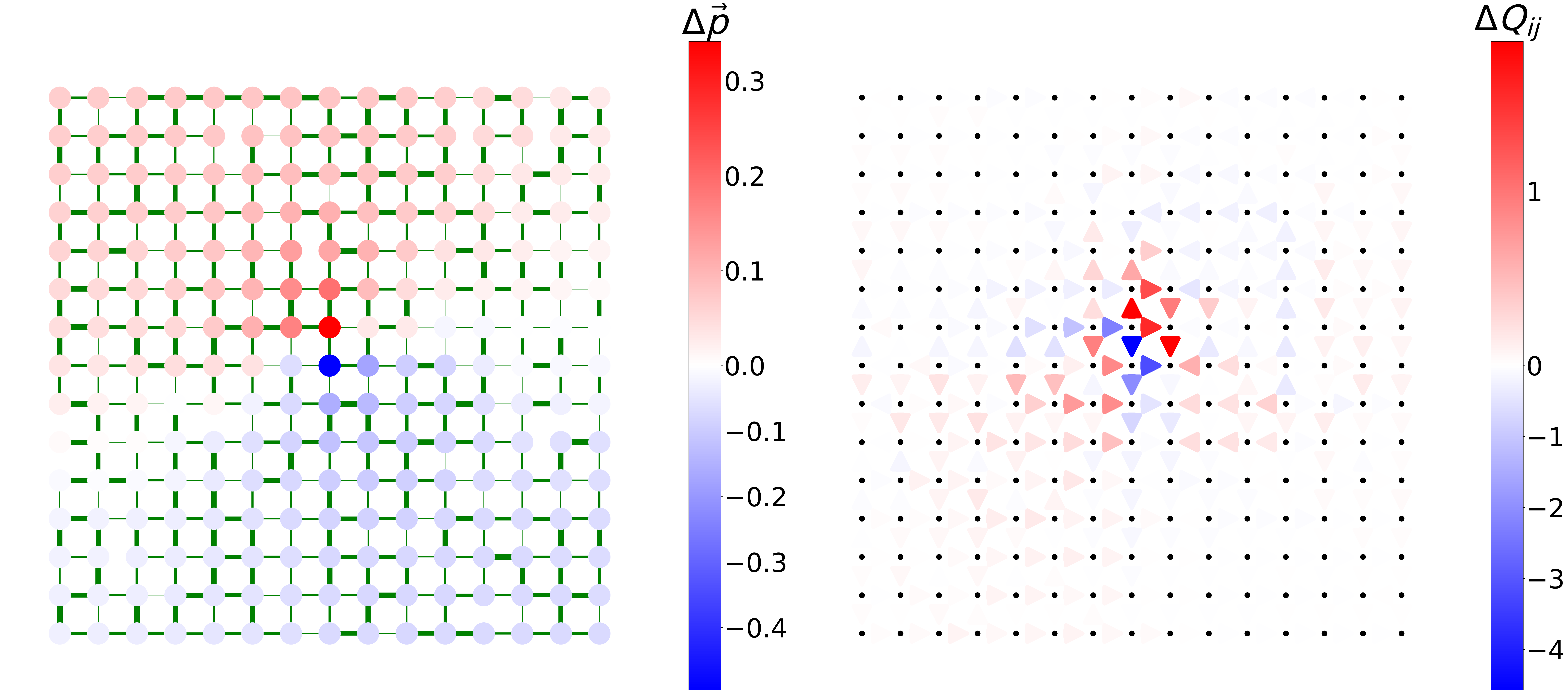}
    \caption{Stationary density states and flux differences after a single-edge failure on a $15 \times 15$ grid network. The left plot shows the changes of the node state after a failure according to the scale on the right, and the arrows in right plot show the corresponding flux difference. The failed edge can be easily localized, being the edge with the maximum flux drop (arrows in the right plot), and it connects the nodes with the maximum density difference (dipole effect). Indeed the state of the node located upstream increases, whereas the state of the node placed downstream decreases.}
    \label{fig: grid diff}
\end{figure*}

\section{Single-edge failure}\label{sec:failure}

We consider the effect of a single-edge failure occurring when a transport network is in the stationary state. From the physics of traffic dynamics point of view, we are assuming that the failure is a consequence of a local congestion whose onset is much faster than the evolution time scales of the network dynamics. Then the failure effect can be modeled by a sudden decrease of an edge weight $w_{kh}$ 
\begin{equation}\label{eq:failure_edge_weight}
    w_{kh} \longrightarrow w'_{kh} = \alpha w_{kh} \quad \text{with} \quad 0 < \alpha < 1.
\end{equation}
and the change of the Laplacian matrix after a link failure is
\begin{equation}
    L \longrightarrow L' = L + \Delta L^{kh}
\end{equation}
with 
\begin{equation}
    \Delta L_{ij}^{kh} = 
    \begin{cases}
        -(1-\alpha)w_{kh} &\text{for } i=j=k \text{ or } i=j=h \\
        (1-\alpha)w_{kh} &\text{for } i=k, j=h \text{ or } i=h, j=k.
    \end{cases}
\end{equation}
After the failure occurs, we have a transient solution 
\begin{equation}
    \vec{p '}(t) = \sum_{\lambda ' \neq 0} c_{\lambda '} e^{-\lambda ' t} \vec{v}_{\lambda '} +  \vec{p^{s'}}
\end{equation}
where $\lambda '$ and $\vec{v}_{\lambda '}$ are respectively the eigenvalues and eigenvectors of the new Laplacian matrix $L'$, and $\vec{p^{s'}}$ is the new stationary solution
\begin{equation}\label{eq:failure}
    0 = -L' \vec{p'} + \vec{s}
\end{equation}
The coefficients $c_{\lambda'}$ are determined by the previous stationary solution $\vec {p^s}$.
If the whole state $\vec{p}(t)$ of the network would be observable, the failure identification problem is trivial. Fig. \ref{fig: grid diff} shows the changes $\Delta \vec{p} = \vec{p^{s'}} - \vec{p^{s}}$ in the stationary solution and the edge flux differences $\Delta Q_{ij} = Q'_{ij} - Q_{ij}$ for a single-edge failure on a $15 \times 15$ lattice network. We note that the edge with the maximum drop flux drop and the nodes with in the maximum density difference (\textit{dipole} effect) localize unambiguously the failure. 
However in real transport networks, real time monitoring all the nodes could be expensive or unfeasible, and, probably, not necessary. We will show that it is possible to perform a dimensionality reduction technique that optimize the localization the single-edge failure, using a limited number of monitored nodes. For this purpose, we apply a clustering method by using the correlation among the nodes dynamics in each cluster.  

%% file: sections/section04.tex
\section{Susceptibility of the network under stochastic perturbation}\label{sec:external noise perturbation}

The susceptibility of the system (\ref{eq:our model}) can be computed by adding to the dynamics (\ref{eq:our model}) a fluctuating term $\vec{\xi}(t)$
\begin{equation}\label{eq:master external noise}
	\dot p_i=\sum_j L_{ij}p_j+s_i + \xi_i(t).
\end{equation}
with the the condition.
$$
\sum_i \xi_i(t)=0 \quad \forall \; t.
$$
to keep the total load fixed. If $\xi(t)$ is a random vector defined by
$$
\xi_i(t)=\eta_i(t)-\frac{1}{N}\sum_k \eta_k(t)
$$
where $\eta_i(t)$ are $N$ independent white noises, eq. (\ref{eq:master external noise}) is a stochastic differential equation. By using the relations
\begin{align*}
	\sum_k E[\eta_i(t) \eta_k(t')] =& \delta(t-t') \sum_k \delta_{ik} = \delta(t-t') \\
	\sum_{kh} E[\eta_k(t)\eta_h(t')] =& \delta(t-t') \sum_{kh} \delta_{kh} = N\delta(t-t')
\end{align*}
the covariance matrix of the random vector $\xi$ reads
\begin{equation}\label{cov0}
	C_{ij}(t'-t) := E[\xi_i(t')\xi_j(t)] = \left ( \delta_{ij}-\frac{1_{ij}}{N}\right ) \delta(t'-t).
\end{equation}
We remark that the  matrix (\ref{cov0}) satisfies
\begin{equation*}
	\sum_j C_{ij} = 0
\end{equation*} 
so that it has the same invariant subspace (\ref{eq:solvability condition}) of the Laplacian matrix $L$. On the invariant subspace (\ref{eq:solvability condition}), the covariance matrix (\ref{cov0}) proportional to the identity matrix. As a consequence the fluctuations $\delta p_i$ of the $i$-node state have no components in the kernel of $L$. We study the perturbation of the stationary state 
\begin{equation}\label{fluct}
	p_i(t)=p^s_i+\delta p_i(t)
\end{equation}
using the master equation (\ref{eq:master external noise}) and we get 
\begin{equation}\label{eq:master external noise fluctuations}
	\delta \dot p_i = -\sum_j L_{ij}\delta p_j + \xi_i(t).
\end{equation}
The equation defines a multidimensional \textit{Ornstein-Uhlenbeck process} \cite{ornstein_U} that can be explicitly solved using in the eigenbasis of $L$
\begin{equation*}
	\delta \vec{p} = \sum_{\lambda} \delta p_\lambda \vec{v}_\lambda \quad \text{and} \quad \vec{\xi} = \sum_{\lambda} \xi_\lambda \vec{v}_\lambda
\end{equation*}
Eq. (\ref{eq:master external noise fluctuations}) reduces
\begin{equation*}
	\begin{aligned}
		\delta \dot p_\lambda &= -\lambda \delta p_\lambda + \xi_\lambda \quad && \text{for} \quad \lambda \neq 0 \\
		\delta \dot p_{0} &= \xi_{0} = 0 \quad && \text{for} \quad \lambda = \lambda_0 = 0
	\end{aligned}
\end{equation*}
and the covariance matrix can be written in the form (see Appendix \ref{sec:appendixA})
\begin{equation}\label{eq: covariance nodes perturbation}
	E[\delta p_i\delta p_j] = \sum_{\lambda\ne 0}\frac{1}{2\lambda}\ket{v^\lambda}_i \prescript{}{j}{\bra{v^\lambda}}.
\end{equation}
which depends only on the spectral properties of the Laplacian matrix (i.e. the structure of the network) and not on the dynamical state of the system.
In the symmetric case, the covariance between two nodes $i$ and $j$ has a positive contribution if the eigenvector $v_\lambda$ does not distinguish between the two nodes (i.e. its components have the same sign for both the nodes), whereas one has a negative contribution when $v_\lambda$ distinguishes the nodes. This result is similar to the idea of spectral clustering with a spring-mass system described in the section\cite{spectral_clustering_with_physical_intuition}. Lower eigenvalues, corresponding to the less energetic oscillations of the network, are associated to eigenvectors whose components have the same sign on large graph subset, corresponding to the large scale structures. Conversely, higher eigenvalues are associated to high energy modes, able to move in opposite directions the nodes connected by a strong link ("dipole" effect). Because of the factor $\lambda^{-1}$, higher eigenvectors contribute less to the covariance matrix. The correlation matrix reads
$$
C[\delta p_i \delta p_j] = \frac{E[\delta p_i\delta p_j]}{\sqrt{E[\delta p_i^2] E[\delta p_j^2] }}
$$
and it defines the linear dependence among the node dynamics. 
\par\noindent
To study dynamical effects of fluctuations, we generalize the previous approach by computing the network sensitivity when the link weights are stochastically perturbed. In this way we can study the dynamical effects of changes in the link weights, that alter the transportation efficiency. 
 At this purpose, we introduce a perturbation $\delta L$ in the Laplacian matrix
\begin{equation}\label{eq:laplacian perturbation}
    \dot{\vec{p}}=-(L+\delta L) \vec{p} + \vec{s}
\end{equation}
with the condition
\begin{equation*}
    \sum_i \delta L_{ij} = 0 \quad \text{and} \quad \delta L_{ij} = \delta L_{ji}
\end{equation*}
Therefore, the coefficients $\delta L_{ij}$ contain $n(n-1)/2$ independent random variables. Recalling the definition  of the Laplacian matrix $L_{i j} = \delta_{i j}d_{j} - w_{i j}$, we have
\begin{equation*}
    \delta L_{ij}(t) = \left[ w_{ij} \delta w_{ij}(t) - \delta_{ij} \sum_{k} w_{ki} \delta w_{ki}(t)\right].
\end{equation*}
We remark that each fluctuation $\delta L_{i j}$ is proportional to the unperturbed weights $w_{i j}$, so that the entries of $L_{ij}=0$ are unaffected. We further require that the random variables $\delta w_{i j}$ have zero average and covariance matrix
\begin{gather*}
    E[\delta w_{ij}(t) \delta w_{kl}(t')] = (\delta_{ik}\delta_{jl}+\delta_{il}\delta_{jk})\delta(t-t')
\end{gather*}
Then the Laplacian fluctuations $\delta L$ is \textit{white noise} random matrix
\begin{equation}\label{eq:laplacian white noise}
    E[\delta L_{kl}(t) \delta L_{hm}(t')] = C_{klhm} \delta(t-t')
\end{equation}
where $C_{klhm}$ is the correlation tensor among the entries of $\delta L_{kl}$.
A perturbation approach to the solution requires to compute the spectral properties of the matrix $L+\delta L$ using the results of the \textit{Random Matrix Theory}\cite{tao_1970}. However to evaluate $E[\delta p_i \delta p_j ]$ in the white noise limit, we can proceed in a more straightforward way by using a fixed point principle. The final expression is computed in Appendix \ref{sec:appendixB}:
\begin{equation}\label{eq:covariance edges perturbation}
    E[\delta p_i\delta p_j] = \sum_{\lambda,\mu\ne 0} \frac{1}{\lambda+\mu} \ket{v^\lambda}_i \left[ \sum_{k,h} \prescript{}{k}{\braket{v^\lambda |C_{kh}| v^\mu}}_h \right] \prescript{}{j}{\bra{v^\mu}}
\end{equation}
where $C_{kh}$ is defined as
\begin{equation}\label{eq: metric matrix}
    C_{kh} = \left\{\begin{array}{ll}
        - [w_{kh}(p_k-p_h)]^2 & \text{for} \quad i \neq j \\
        \sum_i[w_{ki}(p_k-p_i)]^2  & \text{for} \quad i=j
    \end{array}\right.
\end{equation}
and it contains information on the stationary fluxes of the transport network. The term in the squared brackets of eq. (\ref{eq:covariance edges perturbation})
\begin{equation}\label{eq:scalar product}
    S^{\lambda\mu} := \sum_{k,h} \prescript{}{k}{\braket{v^\lambda |C_{kh}| v^\mu}}_h 
\end{equation}
is the scalar product between the eigenvector $v_{\lambda}$ and $v_{\mu}$, defined by the metric matrix $C_{kh}$. Note that $C_{kh}$ has the structure of a Laplacian matrix and, being $C_{kh}$ symmetric and diagonally dominant, it is semipositive-definite. $S^{\lambda\mu} \geq 0$ and it determines how much a given eigenvector couple $\lambda,\mu$ contributes to the covariance matrix (\ref{eq:covariance edges perturbation}). Expanding the scalar product (\ref{eq:scalar product}) we get
\begin{equation*}
\begin{aligned}
    S^{\lambda\mu} = & \sum_{k} 
     \prescript{}{k}{\braket{v^\lambda |C_{kk}| v^\mu}}_h  +
     \sum_{h} \prescript{}{h}{\braket{v^\lambda |C_{hh}| v^\mu}}_h + \\
     & \sum_{k \neq h} \prescript{}{k}{\braket{v^\lambda |C_{kh}| v^\mu}}_h 
\end{aligned}
\end{equation*}
%
%
The biggest contributions to the $S^{\lambda\mu}$ (and therefore to the covariance matrix) are given by the Laplacian eigenvectors $v^\lambda$ that localize on the nodes with higher total incoming and outgoing fluxes (diagonal elements of $C_{kh}$). Moreover, there are negative contributions given by the eigenvectors that represent dipole effect on the links with higher fluxes (off-diagonal elements of $C_{kh}$). This fact will be of great importance for the subsequent discussion. 
As final remark, if the fluxes are all equal
\begin{equation}
    w_{ij} (p_i-p_j) = \frac{1}{\sqrt N} \quad \forall (i,j)
\end{equation}
and the network is complete (there is an edge between all couples of nodes $(i,j)$) we have that 
\begin{equation}
    C_{kh} =
    \begin{cases}
        \sum_i [w_{ki} (p_k-p_i)]^2 = 1-\frac{1}{N} &\text{ if } k=h \\
        - [w_{kh} (p_k-p_h)]^2 = -\frac{1}{N} &\text{ if } k\neq h
    \end{cases}
\end{equation}
Then the metric matrix reads
\begin{equation}
    C_{kh} = \left(\delta_{kh}-\frac{1_{kh}}{N}\right) 
\end{equation}
and consequently
\begin{equation}
    \begin{aligned}
        S^{\lambda\mu} &:= \sum_{k,h} \prescript{}{k}{\braket{v^\lambda |C_{kh}| v^\mu}}_h \\
        &= \sum_{k,h} \prescript{}{k}{\braket{v^\lambda |\delta_{kh}| v^\mu}}_h - \frac{1}{n}\sum_{k,h} \prescript{}{k}{\braket{v^\lambda |1_{kh}| v^\mu}}_h \\
        &= \delta^{\lambda\mu}
    \end{aligned}
\end{equation}
since the eigenvectors are orthogonal and their components sum to $1$ (we are not considering the kernel here). From (\ref{eq:covariance edges perturbation}) we recover the result (\ref{eq: covariance nodes perturbation}).

\section{Dynamics-based partitioning}\label{sec:dynamic-based-partitioning}
A dynamics-based clustering method aims to group together nodes that whose dynamics is highly correlated. Concerning the failure identification problem, an effective dynamic clustering, allows to select \textit{sensor} nodes for each cluster, whose dynamics is representative of the dynamics of the other nodes in the cluster when a failure appears. The covariance matrix (\ref{eq:covariance edges perturbation}) mimics the average single-edge failure behaviour of the network. There are multiple ways to exploit the covariance matrix (\ref{eq:covariance edges perturbation}), or the corresponding correlation matrix, in order to partition the network. In order to observe the network state using a subset of its nodes, we need a clustering method that provides not only communities of nodes but also a representative node for each cluster. 
 At this purpose, we performed a Principal Component Analysis (PCA)\cite{hand_data_2008} of the covariance matrix (\ref{eq:covariance edges perturbation}) to exploit the eigenvectors corresponding to the highest eigenvalues, and to provide an embedding space on which the node states can be projected. We observe that the eigenvectors of the covariance matrix (\ref{eq:covariance edges perturbation}) with highest eigenvalues are localized on the edges with highest fluxes: i.e. these eigenvectors describe the \textit{dipole} effect on the edges with higher fluxes and they are excited when a failure on one of these edges occurs. 
 %
 %
 Furthermore, to select a single sensor node for each cluster, a k-medoid clustering is performed over the nodes coordinates given by the Principal Components embedding. A k-medoid algorithm, differently from the k-means algorithms, provides automatically the representative point for each cluster. More precisely, given the covariance matrix eigenvalues $0\equiv\mu_0 \leq \mu_1 \leq \dots \leq \mu_{N-1}$ and the corresponding eigenvectors $\vec{u}_0,\vec{u}_1 \dots \vec{u}_{N-1}$ the chosen nodes embedding is given by 
 \begin{equation}\label{eq:embedding}
     \left[\sqrt{\mu_{N-1-\tilde k}} \vec{u}_{N-1-\tilde k}, \dots,  \sqrt{\mu_{N-1}} \vec{u}_{N-1} \right]^T
 \end{equation}
where $\tilde k$ is the number of eigenvectors considered for the embedding. The $i-th$ row of (\ref{eq:embedding}) provides the embedding for node $i$, i.e. the k-medoid clustering is applied over the rows of (\ref{eq:embedding}), like in the spectral clustering. We choose the number of eigenvectors used for the embedding to be equal to the number of input clusters for the k-medoid algorithm ($\tilde k = k$). One may use the correlation matrix instead of the covariance matrix to perform the clustering, but the efficiency in localizing edge failures of links with large fluxes using the covariance matrix turns out to be more efficient (see section \ref{sec:results}).\par\noindent
The covariance matrix obtained from the nodes perturbation (\ref{eq: covariance nodes perturbation}) is considered by several papers\cite{prediction-of-allosteric-sites}\cite{edge-based-elastic} and it corresponds to the (pseudo)inverse of the Laplacian matrix $L^+$, so that it has the same spectral properties. A clustering algorithm based on the highest eigenvectors of (\ref{eq: covariance nodes perturbation}) would give similar results as the spectral embedding with k-means, a common structure-based partitioning method. The same relation occurs between the correlation matrix and the normalized spectral embedding with k-means clustering. Conversely, the eigenvectors corresponding to the lower part of the spectrum, multiplied by the square root of the eigenvalues, provides the \textit{Commute Time Embedding} and the related concept of \textit{Resistance Distance}\cite{von_luxburg_tutorial_2007}.

%% file: sections/section05.tex
\section{Failure identification procedure}\label{sec:failure identification}

\begin{figure*}[ht]
    \includegraphics[width=\textwidth]{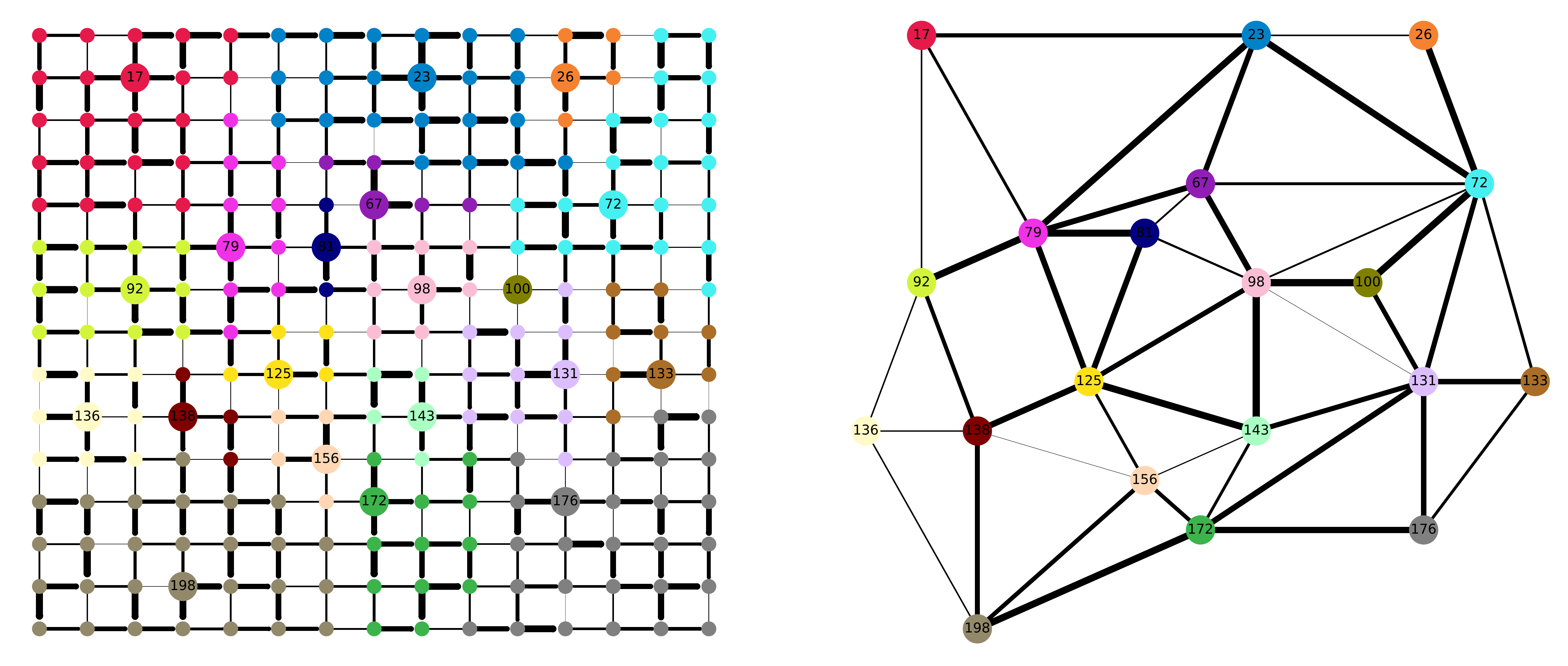}
    \caption{Dimensionality reduction of a $15\times 15$ grid network, from the initial $N=225$ nodes to $20$ nodes. Each sensor node of the reduced graph represents the nodes belonging to a given cluster in the original network. A single edge is set between two sensor nodes if there were edges between the corresponding represented clusters. In the original network, each color denotes a different clusters and the larger nodes are the sensor nodes. The lines thickness is proportional to the edge weights.}
    \label{fig:grid reduction}
\end{figure*}
We assume that the state of the network can be only partially observed using a subset of nodes $S\in\mathcal{N(\mathcal{G})}$, called  \textit{sensor} nodes. The failure identification problem requires first, to define a clustering method to partition the graph, and secondly, to identify a single sensor node for each cluster, that is representative of the state of all the cluster nodes. The aim is to find an optimal partitioning of the network and an optimal sensor placement which maximizes the probability to localize the failure. \par\noindent
Once a partitioning is established and a sensor node is chosen for each cluster, a dimensionality reduction of the network is performed, collapsing each cluster to the sensor node. A \textit{coarse-grained} model of the original graph is built by connecting the sensor nodes when there is at least one link among the nodes of the two represented clusters. The set of links connecting two clusters is called \textit{edge-boundary}. The weights of the links in the coarse-grained graph are the sum of the original link weights belonging to the edge-boundary. This quantity is also called \textit{cut size}. Let $N$ be the size of the original network and $r$ the number of sensors set, i.e. the size of the reduced network, we define by $C\in\mathbb{R}^{N\times r}$ the indicator matrix for the clusters
\begin{equation}\label{eq:indicator matrix}
    C_{ij} = 
    \begin{cases}
    1 \quad \text{if node $j$ belongs to cluster $i$}  \\
    0 \quad \text{otherwise}
    \end{cases}
\end{equation}
and therefore the \textit{coarse-grained Laplacian} of the reduced network reads
\begin{equation}\label{eq:reduced_laplacian}
    L^{R} = C^T L C.
\end{equation}
Correspondingly, the external forcing on the reduced network is computed by summing the forcing over the nodes belonging to the same cluster
\begin{equation}\label{eq:forcing reduced}
    \Vec{s}^{\;R} = C^T \vec{s}.
\end{equation}
The master equation for the diffusion dynamics on the reduced network reads
\begin{equation}\label{eq:reduced network linear model}
    L^{\;R} \vec{p}^{\;R} = \Vec{s}^{\;R}
\end{equation}
where $\vec{p}^{R}$ is the stationary state of the reduced network. Relation between the properties of the spectrum of $L^R$ and $L$ are discussed in Appendix \ref{sec:appendixC}. Fig. \ref{fig:grid reduction} shows an example of graph reduction for a grid network with $N=225$ that is reduced to $r=20$.\par
\begin{figure*}[!t]
    \includegraphics[width=\textwidth]{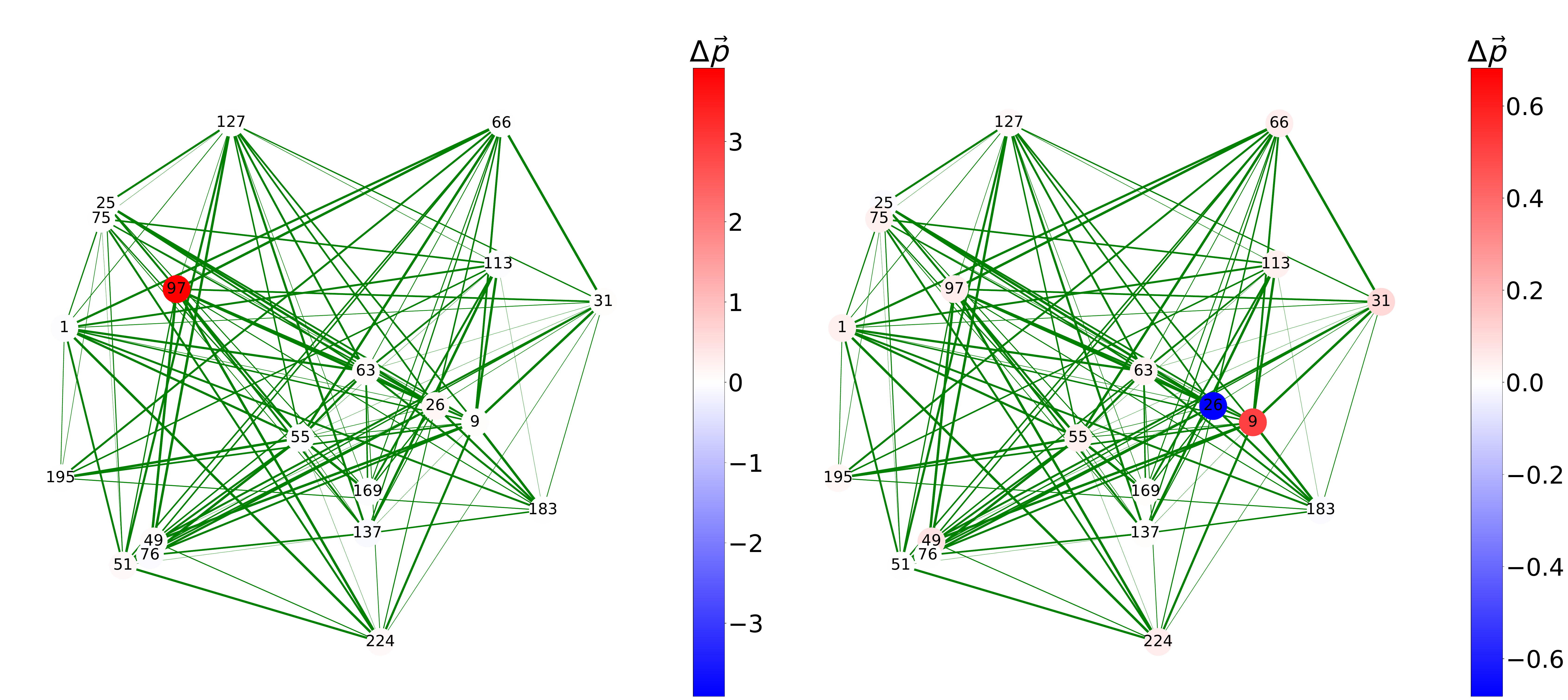}
    \caption{Expected behavior for an edge failing inside a single cluster (left) and between two clusters (right). In the former case we observe a monopole effect, whereas in the latter case one has a dipole effect. In both cases, we are able to identify the group of edges to which the failed one belongs.}
    \label{fig: monopole dipole}
\end{figure*}

\noindent
To localize a single-edge failure observing the signal from the sensor nodes, we have to consider two different situations. If the failure is internal to a cluster, we expect to detect a relevant density change in the representative sensor node, and a considerably lower response from the other sensor nodes in the network. We call \textit{monopole} the signal pattern corresponding to this case. Conversely, if the failed link belongs to the edge boundary of two clusters, we expect to see a net \textit{dipole} effect between these two clusters. These ideal cases are illustrated in Fig. \ref{fig: monopole dipole}.
In other cases we have a signal pattern that involves the whole network. Nevertheless, if there is a net separation among the regions monitored by sensors with a positive and negative signal difference, it is still be possible to identify the location of the failure. In particular, the failed edge should be localized around the sensor nodes corresponding to the maximum drop in the signal difference. This is in agreement with the situations shown in Fig. \ref{fig: grid diff} applied to the coarse-grained network. Then, after an edge failure is detected and the system relaxes to the new stationary state, one first selects the couple of adjacent (in the coarse-grained graph) sensor nodes with the highest density drop (i.e. the highest observed dipole). There are four different possible failures. The failed link is internal to the cluster with a positive density difference (positive pole), or with a negative density difference (negative pole). Alternatively, the failed link belongs to the edge boundary of the positive and negative poles. Finally, the failed link belongs to none of the above three cases. To choose among these possible situations, we study the effect of any link failure in the original network for each partitioning and sensors distribution, using the coarse-grained network defined by eq. (\ref{eq:reduced_laplacian}). If the failure is internal to a cluster $k$, we expect to see a monopole: i.e.
$$
\Delta \vec{p}_{k}^{\;R} = c (0,\dots,\underbrace{1}_{k},\dots,0)^T.
$$
where $c$ is the monopole strength. Conversely, if the failed edge belongs to the edge-boundary between two clusters, we expect to see a dipole effect. However, the exact ``form" of this dipole is not known a priori. In particular, we are interested in the relative strength between the positive and negative part of the dipole, in order to distinguish the external failures from the internal ones. It is quite improbable that an external failure leads to a ``perfectly balanced" dipole (i.e. the positive and negative parts have the same value but opposites signs). Therefore, we test the failure of each edge on the coarse-grained network and we save the response over its nodes:
$$
\Delta \vec{p}_{e}^{\;R} = \frac{\vec{p}_{e}^{\;R} - \vec{p}^{\;R}}{\|\vec{p}_{e}^{\;R} - \vec{p}^{\;R}\|} \quad\quad \forall e\in\mathcal{E}(\mathcal{\;R})
$$
where $\vec{p}_{e}^{\;R}$ is the new stationary state after the failure of the edge $e$ and $\vec{p}^{\;R}$ is the initial stationary solution (see equation (\ref{eq:reduced network linear model})). The relative dipole strength is defined by the ratio between the positive pole response and the negative one
\begin{equation}\label{eq:relative dipole strength}
    r = \frac{\Delta p_{e,+}^{\;R}}{\Delta p_{e,-}^{\;R}}.
\end{equation}
Since every edge $e$ of the coarse-grained network is the sum of the links between two clusters on the original network (the edge-boundary; cfr. eq. (\ref{eq:reduced_laplacian})), $\Delta \vec{p}_{e}^{\;R}$ and the relative dipole strength (\ref{eq:relative dipole strength}) should reflect the average signal we would measure after a failure of one the represented links in the original network. We remark that this procedure can be performed ``off-line", after the definition of the coarse-grained network. 
We save the set of expected signals given by
$$
\Delta \vec{p}_{t}^{\;R} = \Delta \vec{p}_{i}^{\;R}  \cup \Delta \vec{p}_{e}^{\;R}
$$ 
where $i,e,t$ are indices that span over the failure cases and stands for ``internal", ``external" and ``total"; $t$ denotes all the possible failures observed on the coarse-grained network.
Once a link failure occurs in the initial network, we measure a difference in the stationary solution over the sensor nodes $\Delta \vec{p}_{|_{\mathcal{S}}}$. We remark that the signal obtained from the sensor nodes does not correspond, in general, to the signal of the coarse-grained network $\Delta \vec{p}_{t}^{R}$. But our ansatz for the detection of the failed cluster is to select the one whose response is more similar to $\Delta \vec{p}_{|_{\mathcal{S}}}$ among the recorded $\Delta \vec{p}_{t}^{\;R}$. We introduce a similarity measure using the \textit{cosine similarity} (\ref{eq:cosine}), i.e. the scalar product of normalized vectors. We take the normalized vector so that it does not depend on the value of $\alpha$ in eq. (\ref{eq:failure_edge_weight}). Therefore the similarity values read
\begin{equation}\label{eq:cosine}
s_t = \frac{|\Delta \vec{p}_{|_{\mathcal{S}}} \cdot \Delta \vec{p}_{t}^{\;R}|}{|\Delta \vec{p}_{|_{\mathcal{S}}} | | \Delta \vec{p}_{t}^{\;R}|}
\end{equation}
with maximum values 
$$
\tilde s_t = \operatorname{max}_{t} (s_t)
$$
and hence our guessed for the failure detection is
$$
\tilde t = \operatorname{argmax}_{t} (s_t).
$$
\begin{figure}
    \begin{tikzpicture}
        \draw[<->] (-3,0)--(3,0) node[right] {\textbf{Positive pole}};
        \draw[<->] (0,-3)--(0,3) node[above]{\textbf{Negative pole}};
        \draw[line width=2pt,blue,-stealth](0,0)--(0,-2) node[anchor=south west]{}; 
        \draw[line width=2pt,red,-stealth](0,0)--(2,0) node[anchor=north east]{};
        \draw[line width=2pt,green,-stealth](0,0)--(1.732,-1) node[anchor=west]{Expected dipole signal};
        \draw[line width=2pt,black,-stealth](0,0)--(1.414,-1.414) node[anchor=west]{Observed signal};
        \draw[black] (0,0) circle (2cm);
        \node[black,below] at (2.2,0) {1};
    \end{tikzpicture}
    \caption{Expected dipole signal (from the ``off line'' breaking tests on the coarse-grain network) and measured signal (on the sensor nodes of the real network after a single edge failure) projected on the subspace of the maximum observed dipole for better visualization (i.e. we compute the relative dipole strength of eq. (\ref{eq:relative dipole strength})). In the considered situation, the measured signal is closest to the signal expect from a failure between the node i and node j.}
    \label{fig:circle}
\end{figure}
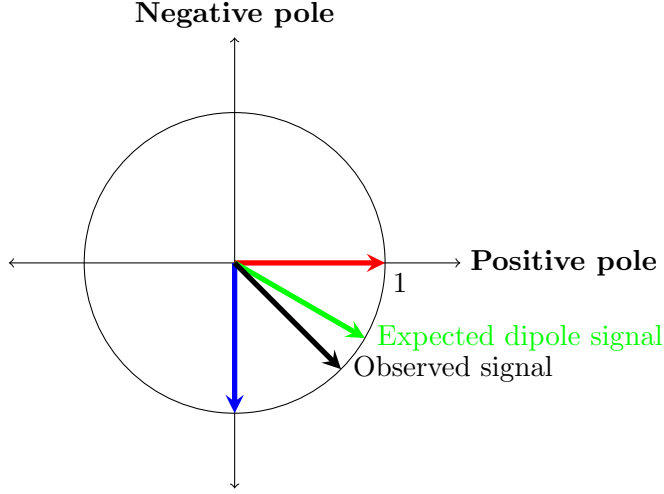
Fig. \ref{fig:circle} gives a visualization of the proposed procedure. The higher is the similarity between the measured signal and the recorded ones, the more probable is the correct localization of the cluster containing that particular failure. As a benchmark value to measure the failure localization efficiency of a particular partitioning, we use the cosine similarity value itself if our guessed cluster is the correct one, and we set the value to zero if it is the wrong one. However, since the bigger is a cluster and the easier is to localize a failure (a single cluster that spans the whole network always localizes the failure), we divide the cosine similarity by the total number of internal links $M_t$ and we define an efficiency value
\begin{equation}\label{eq:efficiency value}
\text{Efficiency value} = \frac{100}{M}\sum_t b_t
\end{equation}
where $M$ is the total number of links in the network and $b_t$ is the benchmark value of the $t$ cluster defined by
\begin{equation*}
    b_t = 
    \begin{cases}
        \tilde s_t/M_t \quad & \text{ if correct } \\
        0 \quad & \text{ otherwise.}
    \end{cases} 
\end{equation*}
The above procedure is performed for each failure of the original network. Note that the off-line procedure has to be performed only once for a sensors distribution. The average over all the benchmark values is the value plotted for the $y$-axis in all the figures of the next section. When the clusters number tends to the number of nodes $N$, the detection procedure classifies correctly each failure and the efficiency value (\ref{eq:efficiency value}) tends to the $100\%$ percentage. 
\begin{figure*}[ht]
    \centering
    \begin{subfigure}
       \centering
        \includegraphics[width=.48\linewidth]{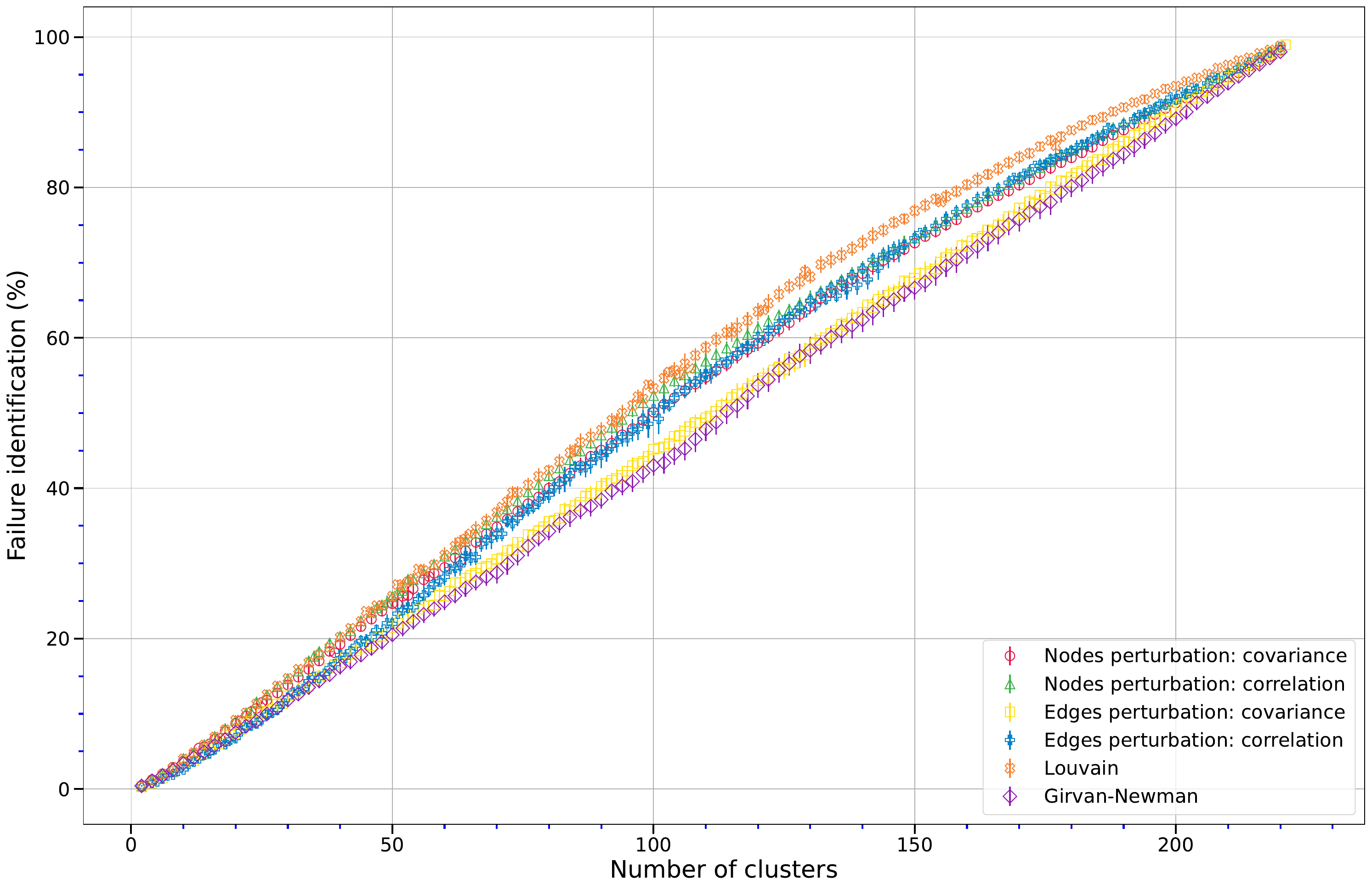}
    \end{subfigure}
    \begin{subfigure}
        \centering
        \includegraphics[width=.48\linewidth]{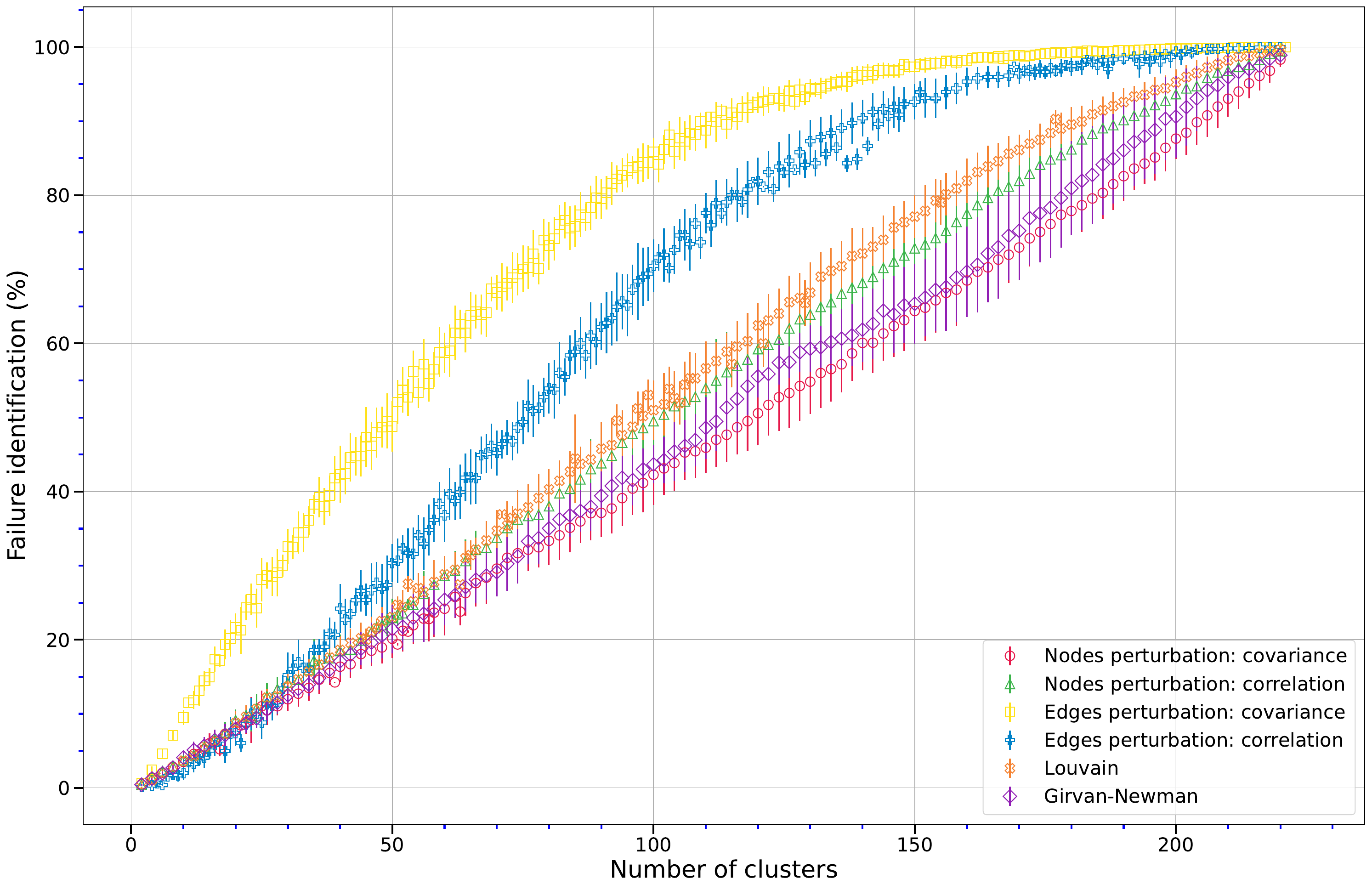}
    \end{subfigure}
    
    \medskip
    
    \begin{subfigure}
        \centering
        \includegraphics[width=.48\linewidth]{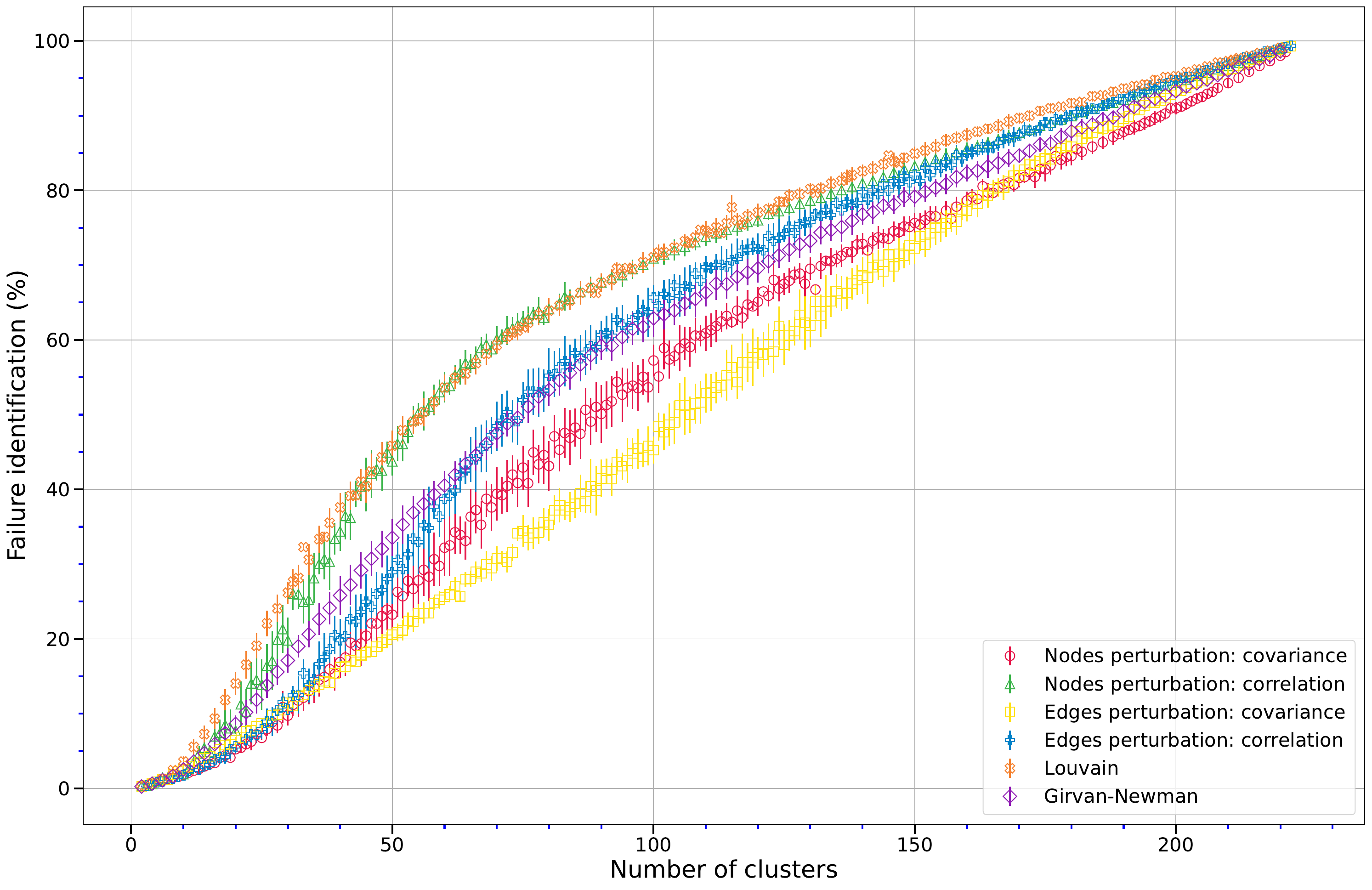}
    \end{subfigure}
    \begin{subfigure}
        \centering
        \includegraphics[width=.48\linewidth]{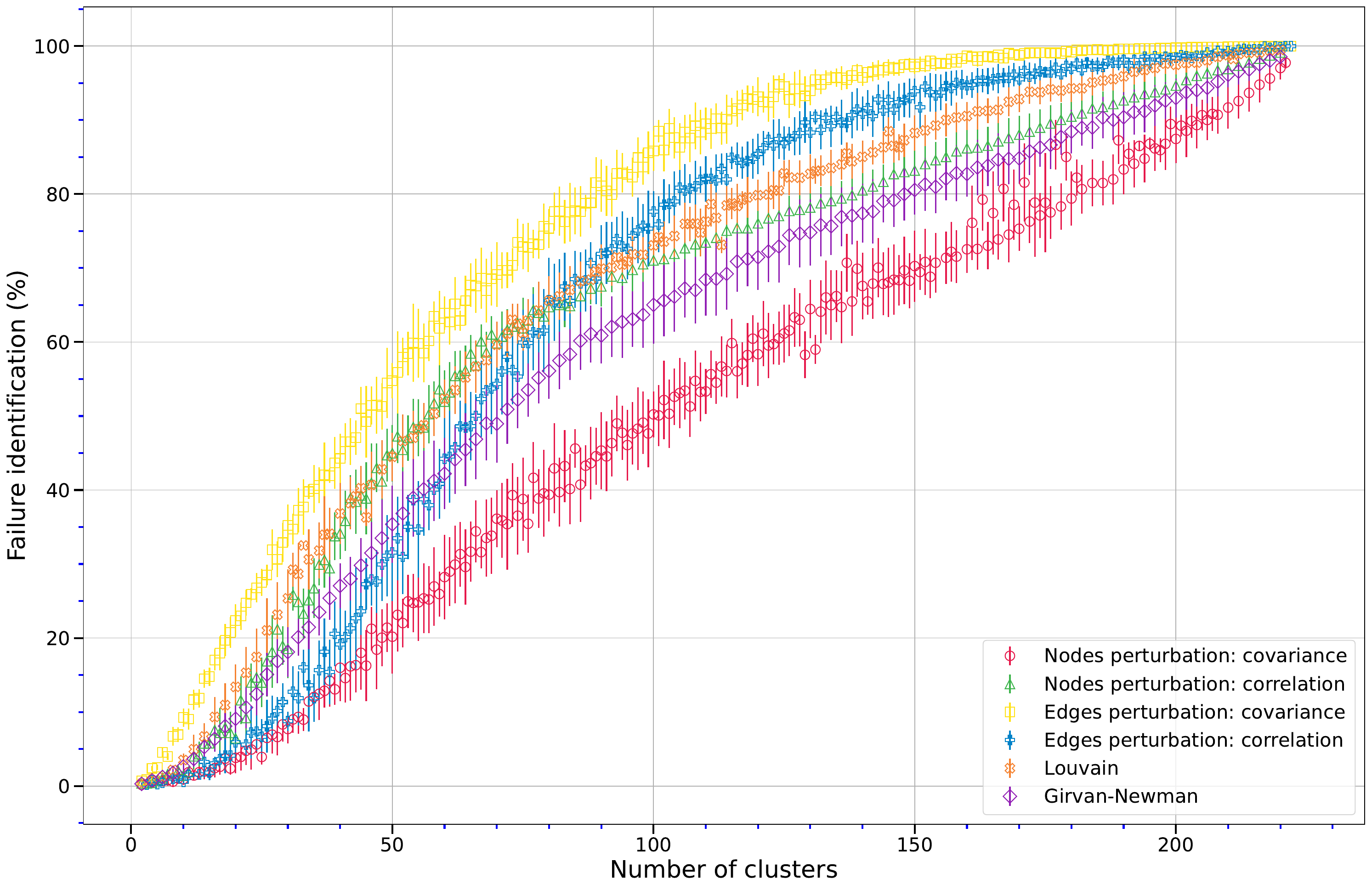}
    \end{subfigure}
    \caption{Failure detections efficiency according to eq. (\ref{eq:efficiency value}) as a function of clusters number for grid networks (top) and Erdos-Renyi random networks (bottom) with average connectivity $<k>= 4$. The number of nodes is $N=225$ for both cases and we average over 20 realizations of the network structure reporting the standard deviation on each point.  In the Left figures we show detection efficiency over all the possible single-edge failures. In the Right pictures we show detection efficiency over the edges, whose flow value is more than the $75$-th percentile of the flow distribution. Different symbols correspond to different clustering algorithms as reported in the pictures. \textcolor{red}{As discussed in the main text, the edge covariance has a higher efficiency for failures involving higher fluxes, at the expense of the lower fluxes.
    }
    }
    \label{fig:failures results}
\end{figure*}

%% file: sections/section06.tex
\section{Failure Detection Results}\label{sec:results}

To evaluate the efficiency of the failure detection procedure we have applied the clustering algorithm to the different graph structures. We present the cases of weighted grid networks and Erdos-Renyi random networks using both the covariance and the correlation matrices obtained by the node stochastic perturbation (\ref{eq: covariance nodes perturbation}) and the link stochastic perturbation (\ref{eq:covariance edges perturbation}). The analogous results for the scale-free and regular networks are reported in the Supplementary Material. These network typologies may represent the spatial structure of transport networks, where congestion phenomena are particularly relevant. 
We compare the proposed failure detection procedure with other approaches based on traditional clustering algorithms based on the network topology.
We first considered the \textit{modularity maximization} clustering method\cite{G-N_modularity}, with the Louvain method. The second clustering method was the divisive algorithm by Girvan and Newman\cite{G-N_algorithm}, that detects communities by removing edges iteratively from the original graph until the underlying community structure of the network emerges. For both algorithms we used the versions implemented in \textit{Networkx}\cite{networkx}, a Python package for the study both of structural and dynamical complex networks. 
In Fig. \ref{fig:failures results} we show the benchmark efficiency value defined by eq. (\ref{eq:efficiency value}) varying the number of clusters for the Grid network and the Erdos Renyi random networks. The number of nodes for both the networks is $N=225$ and the average degree is $k=4$. All the results are averaged over $20$ realizations of the network, with weights following a uniform probability distribution $w_{ij} \sim U(0,1)$. Also the sinks are taken uniformly distributed $s_{i} \sim U(-1,0)$, except from the sources. There are $5$ sources for each network, whose values satisfy the balance condition (\ref{eq:solvability condition}) to get a stationary solution.
In the left pictures of Fig. \ref{fig:failures results} we compute the detection efficiency (\ref{eq:efficiency value}) considering all the possible failures regardless the flux across the link. In the case of a uniform grid all the methods have a similar performance. Conversely in the case of random networks, we see that a topological clustering as the Louvain method or the clustering based on the correlation matrix (\ref{eq: covariance nodes perturbation}) have a better performance for a number of clusters $\simeq N/4$ than a dynamical clustering. On the right pictures we compute the detection efficiency when the failure happens in the links with a high flux (more than the $75$-th percentile of the flow distribution). In this case for both the grid networks and the random networks the dynamical clustering based on the covariance matrix (\ref{eq:covariance edges perturbation}) performs better than the other methods with a low number of clusters. The improve of the performance is remarkable in the case of the grid networks.
Therefore the proposed partitioning method based on the covariance matrix (\ref{eq:covariance edges perturbation}) shifts the sensibility from failures involving lower fluxes to higher fluxes. The dynamic partitioning creates clusters with positive correlated edges within the same cluster, whereas the negative correlated edges lie in the edge boundary between two clusters. These links contains generally the higher fluxes since the proposed failure identification algorithm is more sensitive on failures occurring between two clusters. 

\textcolor{red}{As discussed in Section \ref{sec:failure identification}, the failure detection algorithm assigns higher scores to clustering methods that, for a given failed edge, can clearly identify the associated dipole or monopole. We remark that no sensor placement criterion can achieve high detection rates for all edge failures; this would only be possible by monitoring every node in the network (all plots in Fig. \ref{fig:failures results} reach 100\% failure identification as the number of sensors increases). In the possible applications of our approach the advantage is to optimize the detection efficiency on a subset of edges of interest. The proposed clustering (edge perturbation) shifts efficiency from lower-flux edges to higher-flux edges (see Fig. 1 of Supplementary Material), as illustrated in Fig. \ref{fig:failures results}.}

\textcolor{red}{
By contrast, node perturbation relies on the same information as spectral clustering algorithms (see section \ref{sec:dynamic-based-partitioning}), capturing edge weights but not fluxes. Since high-flux edges do not always correspond to high-weight edges, the node covariance method performs worse than the edge covariance method for edges with higher fluxes. We remark that fluxes may be unknown if the network structure or demands  $q_i$ are not available. But, by perturbing edges and recording node responses, covariances can still be computed, enabling the proposed coarse-graining. This approach is not possible with other methods that require full knowledge of the network structure.
}

\textcolor{red}{As a final remark, we note that equation (\ref{eq:covariance edges perturbation}) requires the diagonalization of the Laplacian matrix, whose computational cost may scale as $O(N^3)$. However, typical transport networks size at the coarsening level studied in network science is some thousands of nodes. Diagonalizing matrices of this size is feasible with users laptops. Results with a higher number of nodes are shows in the Supplementary Information.
} 

%% file: sections/section07.tex
\section{Conclusions}\label{sec:conclusions}

Transport systems are ubiquitous in nature and modern societies, displaying several structures and dynamical features. Their topology can be effectively encoded by a network structure, where the nodes entity exchanges particles according to a continuity master equation. Diffusion processes are commonly studied dynamical models for transport networks, whose evolution is defined by the spectral properties of the Laplacian matrix of the underlying graph. In this paper we cope with the problem of failure detection of a network link in a transport system for which a partial observability is available. A link failure may simulate the congestion effect where the physical interactions among the particles reduce abruptly the flux of the link with an impact on the network transport capacity. In this framework we propose a dynamical clustering method based on the susceptibility measure under stochastic perturbations of the network structure to divide the network into connected clusters and to introduce an optimal sensors displacement, which enhance the detection efficiency of failure events, identifying the corresponding cluster. 

\textcolor{red}{Network susceptibility was introduced in \cite{network_susceptibilities} to analyze the global network response to the failure of a single edge or node. Our approach differs by examining the network's response following a perturbation of the entire structure. In contrast to previous works, we leverage the linear response encoded in the covariance matrix (\ref{eq:covariance edges perturbation}) to coarse-grain the network.}

\textcolor{red}{As a final remark, we note that equation (\ref{eq:covariance edges perturbation}) requires the diagonalization of the Laplacian matrix, whose computational cost may scale as $O(N^3)$. However, typical transport networks size at the coarsening level studied in network science is some thousands of nodes. Diagonalizing matrices of this size is feasible with users laptops. Results with a higher number of nodes are shows in the Supplementary Information.
} 

Our method compares the dynamics of the initial network with the dynamics of a coarse-grained model, whose nodes represent the chosen clusters, to identify the cluster or the set of links between two clusters, where the failure happens. We have studied the efficiency of the dynamical clustering method comparing the results with analogous results obtained using well-known clustering methods. The simulations on different network structures show that the proposed clustering method is more efficient in localizing single-edge failures which involve links with high flux. In the case of transport networks these are the links where a congestion effect may appear assuming that the flow dynamics can be described by a flow-density fundamental diagram\cite{daganzo2008}. Therefore, our approach could be useful for monitoring the dynamics of a transport networks when a partial observability is available. Moreover the method provides an algorithm to optimize the sensors placement in the network to maximize the efficiency in the failure detection. These results could be relevant in the applications to the traffic monitoring problem in any transport network. If one considers the time scales of the congestion dynamics\cite{gonzalez2023} it would be possible to develop algorithms for an early detection of the congestion localization and for applying control strategies to mitigate the congestion effects.

%% file: sections/acknowledgements.tex
\section*{Acknowledgements} \label{sec:acknowledgements}
This work is part of the research activity developed by the authors within the framework of the “PNRR”: SPOKE 7 “CCAM, Connected Networks and Smart Infrastructure” - WP4

%% file: sections/appendix.tex
\section{}
\label{sec:appendixA}
The formal solution of (\ref{eq:master external noise fluctuations}) can be written in the form
\begin{equation}\label{external noise solution}
	\delta p_i(t)=\int_0^t e^{-L_{ij}(t-u)}  \xi_j(u)du
\end{equation}
To study the causality relations among the nodes, we compute the covariance matrix restricted to the subspace (\ref{eq:solvability condition}), where $L$ is invertible (fixing $t<s$)
\begin{equation}
E[\delta p_i(t)\delta p_j(s)] = \int_0^t dt e^{- L_{ik}(t-u)} e^{- L^T_{kj}(t-u)} 
\end{equation}
where we use $E[\xi_k(u)\xi_h(s)]=\delta (u-s)\delta_{kh}$. The stationary covariance is achieved taking the limit $t\to\infty$ 
\begin{equation}\label{eq:stationary covariance matrix}
	E[\delta p_i\delta p_j] := \lim_{t\to\infty}E[\delta p_i(t)\delta p_j(t)] = \int_0^\infty dt e^{- L_{ik}t} e^{- L^T_{kj}t}.
\end{equation}
Using Dirac notation, we decompose $L$ is its eigenbase
$$
\hat L=\sum_{\lambda\ne 0} \ket{v^\lambda} \lambda\bra{v^\lambda}
$$
and we have that 
$$
e^{\hat L} = \sum_{\lambda\ne 0} \ket{v^\lambda} e^{\lambda} \bra{v^\lambda}
$$
Eq. (\ref{eq:stationary covariance matrix}) reads
\begin{eqnarray}
	E[\delta p^2] &=& \sum_{\lambda,\mu\ne 0}\ket{v^\lambda}\braket{v^\lambda | \bar v^\mu}\bra{\bar v^\mu}\int_0^\infty e^{-(\lambda+\mu)t}\nonumber \\
	&=&\sum_{\lambda,\mu\ne 0}\frac{1}{\lambda+\mu}\ket{v^\lambda}\braket{v^\lambda |\bar v^\mu}\bra{\bar v^\mu}. \nonumber
\end{eqnarray}
where $\bar v_\lambda$ denotes the dual base. 
In the general case $\braket{v^\lambda | \bar v^\mu}=G^{\lambda\mu}$ is the inverse of the metric matrix in the eigenvector base and in the case of a symmetric matrix (undirected network) we have that $\braket{v^\lambda | \bar v^\mu}=\delta_{\lambda\mu}$ so that the expression reduces to
\begin{equation}
	E[\delta p_i\delta p_j] = \sum_{\lambda\ne 0}\frac{1}{2\lambda}\ket{v^\lambda}_i \prescript{}{j}{\bra{v^\lambda}}.
\end{equation}

\section{}\label{sec:appendixB}
Expanding perturbatively the state variable $\vec{p}$, the equation above can be linearized, turning the multiplicative noise into an additive noise, that is easier to deal with. We perform a perturbation expansion near the stationary state $\vec{p^{s}}$
$$
\vec{p}=\vec{p^{s}}+\delta \vec{p}
$$
and we get 
\begin{equation}\label{eq:linearmasterp}
    \delta \dot p_i=-L_{ij}\delta p_j -\delta L_{ij}p_j^s+O(\delta L \delta \vec{p})
\end{equation}
that is a multidimesnional Orstein-Uhlenbeck equation when $\delta L$ is a white noise. The solution is
\begin{equation}\label{eq:formal solution linearized}
    \delta p_i(t) = -\int_0^t  e^{-L_{ik}(t-u)}  \delta L_{kl}(u)p_l^s du.
\end{equation}
and one can compute the covariance matrix taking into account eq. (\ref{eq:laplacian white noise}) 
\begin{align*}
    &E[\delta p_i\delta p_j](t) =\nonumber \\
    &= \int_0^t du \int_0^t du'  e^{-L_{ik}(t-u)} e^{-L_{jh}(t-u')} C_{klhm} \delta(u-u') p_l^s p_m^s \nonumber \\
    &= \int_0^t du e^{-L_{ik}(t-u)} e^{-L_{jh}(t-u)} C_{klhm} p_l^s p_m^s
\end{align*}
In the limit for $t\to\infty$ we get
\begin{equation}\label{covariance linearized laplacian perturbation}
    E[\delta p_i\delta p_j] = \int_0^\infty  e^{-L_{ik}t} e^{-L_{jh}t}C_{klhm} p_l^s p_m^s dt.
\end{equation}
and we explicitly compute
\begin{equation}\label{eq:C}
    C_{kh} := C_{klhm}p_l^s p_m^s = E[\delta L_{kl}(t) \delta L_{hm}(t)] p_l^s p_m^s
\end{equation}
Considering each entry of the Laplacian perturbation, if $k\neq h$, we have that 
$$
C_{kh} =
\begin{cases}
    -w^2_{kh} p_k^2 & \text{for } l=m=k,\\
    -w^2_{kh} p_h^2 & \text{for } l=m=h, \\
    w^2_{kh} p_k p_h & \text{for } (l=h, m=k) \text{ or } (l=k, m=h), \\
    0 & \text{otherwise}
\end{cases}
$$
and 
$$
C_{kh} =w^2_{kh}(2p_kp_h - p_k^2 - p_h^2) = -[w_{kh}(p_k-p_h)]^2
$$
whereas the diagonal elements are
$$
C_{kk} =
\begin{cases}
    \sum_{i\neq k} (w_{ki} p_i)^2 & \text{for } l=m\neq k ,\\
    p_k^2 \sum_{i\neq k}w_{ki}^2 & \text{for } l=m=k, \\
    -p_k\sum_{i\neq k}w_{ki}^2 p_i & (l=k \text{ and } m\neq k) \\
    & \text{ or } (m=k \text{ and } l\neq k),\\
    0 & \text{otherwise}
\end{cases}
$$
and we get 
$$
C_{kk}=\sum_{i\neq k}w_{ki}^2(p_k^2-2p_k p_i+p_i^2) = \sum_i [w_{ki}(p_k-p_i)]^2
$$
Then the matrix $C_{kh}$ has the Laplacian character
\begin{equation}
    \sum_k C_{hk} = 0 \quad \text{and} \quad C_{hk} = C_{hk}
\end{equation}
according to its definition (\ref{eq:C}). The matrix $C_{kh}$ gives information on the variations of the effect of perturbation throughout the network when we are in a stationary state. The covariance of the fluctuations (\ref{covariance linearized laplacian perturbation}) can be computed by representing the Laplacian in its eigenbasis
$$
\hat L = \sum_{\lambda\ne 0} \ket{v^\lambda} \lambda\bra{v^\lambda} \quad \hat L^T = \sum_{\lambda\ne 0} \ket{\bar v^\lambda} \lambda\bra{\bar v^\lambda}
$$
We have the expression
\begin{equation}
    \label{covp}
    E[\delta p_i\delta p_j] = \int_0^\infty \sum_{\lambda,\mu\ne 0} \ket{v^\lambda}_i e^{-\lambda t} \prescript{}{k}{\braket{v^\lambda | C_{kh} | \bar v^\mu}}_h e^{-\mu t} \prescript{}{j}{\bra{\bar v^\mu}} 
\end{equation}
where we can solve directly the time-dependent part with
$$
\int_0^\infty e^{-(\lambda+\mu)t} = \frac{1}{\lambda+\mu}
$$
Finally eq. (\ref{covp}) reads
\begin{equation}
    E[\delta p_i\delta p_j] = \sum_{\lambda,\mu\ne 0} \frac{1}{\lambda+\mu} \ket{v^\lambda}_i \left[ \sum_{k,h} \prescript{}{k}{\braket{v^\lambda |C_{kh}| v^\mu}}_h \right] \prescript{}{j}{\bra{v^\mu}}
\end{equation}
where we have exploited the symmetry of the Laplacian.

\section{}\label{sec:appendixC}
The eigenvalues of a symmetric Laplacian matrix can be interpreted as the energies of the normal modes of a spring-mass system. In this analogy, the nodes of the network are the masses and the edges corresponds to the springs, with elastic constant equal to the edges weights $w_{ij}$. If the masses are set $=1$, the system dynamics is governed by the unnormalized Laplacian matrix, with equation 
\begin{equation}\label{eq:spring-mass eom}
	\ddot{x_k} = -\sum_j L_{kj}x_j
\end{equation}
Lower eigenmodes represent oscillations that spread through the whole network, like the Fiedler eigenvector. Higher eigenmodes correspond to oscillations of high energy, and they are localized on the edges with higher weights \cite{localization_laplacian}. These are the only eigenmodes able to to generate a dipole change on strongly connected nodes, i.e. to distinguish them from the other. In the diffusion dynamics, the higher eigenmodes, that involve strongly connected nodes, correspond to the fast relaxing transient states and these nodes are   perceived as a single 'particle' by the lower eigenmodes. 
According to this remark,  we develop a procedure to glue strongly connected nodes to reduce the graph dimension. This means that the weight of their connecting links diverge, but the state of the nodes, seen by the lower eigenmodes, is almost invariant. Moreover, the corresponding eigenvalue tends to infinity.
We iterate this procedure by gluing the couples of the most connected nodes. From a mathematical point of view,  the collapsing procedure means to perform a limit $\lambda \to \infty$ of the eigenvalues, associated to the eigenvectors that distinguish the strongly connected nodes. The idea is that the other eigenvectors are slightly modified by this procedure and the reduced network could give relevant information on the the whole graph, since each node also represents the other nodes collapsed to it.  This reduced version of the graph is called \textit{coarse-grained graph}. If in the reduced network there are nodes belonging to the original one, this procedure is also called network \textit{down-sampling} \cite{shuman_emerging_2013}.
To study how the spectral properties of the Laplacian matrix changes during the reduction procedure is a key issue to control the relation between the initial graph and the reduced one. For a given link, if its weight $w_{kh}$ is increased by
\begin{equation}\label{eq:edge_increase}
	w_{kh} \longrightarrow w'_{kh} =  w_{kh} + \gamma \quad \text{with} \quad \gamma > 1.
\end{equation}
the corresponding Laplacian matrix reads
\begin{equation*}
	L \longrightarrow L' = L + \gamma \delta L
\end{equation*}
with 
\begin{equation*}
	\delta L_{ij} = 
	\begin{cases}
		\frac{1}{2} &\text{for } i=j=k \text{ or } i=j=h \\
		-\frac{1}{2} &\text{for } i=k, j=h \text{ or } i=h, j=k.
	\end{cases}
\end{equation*}
$\delta L$ is a Laplacian matrix an eigenvalue $\lambda=1$ corresponding to the eigenvector 
\begin{equation*}
		\lambda = 1 \implies \vec{v} = \left( 0,\dots, 0, \underbrace{1/\sqrt 2}_k, 0, \dots, 0, \underbrace{-1/\sqrt 2}_h,0,\dots,0\right)^T 
\end{equation*}
and for the invariant subspace associated to the null eigenvalue, a possible basis choice might be
\begin{equation}\label{eq:lambda 0}
	\vec{u} =
	\begin{cases}
		\left(1,0,\dots,0 \right)^T \\
		\left(0,1,\dots,0 \right)^T \\
		\qquad\quad \vdots \\
		\left(0,\dots, 0, \underbrace{1/\sqrt 2}_k, 0, \dots,0, \underbrace{1/\sqrt 2}_h, 0, \dots, 0 \right)^T
	\end{cases}.
\end{equation}
These eigenvectors can be arranged row-wise in a matrix $\Pi^{\perp}$ that is the projector over the subspace orthogonal to the dipole. 
We call the eigenvector $\vec{v}$ a \textit{dipole} vector, since it represents the above mentioned high energy eigenmode able to distinguish the two strongly connected nodes. The matrix $\delta L_{ij} \equiv \Pi_{i j}^{\|}$ is a projector on the subspace generated by the dipole vector $\vec{v}$. To study the spectral properties of the perturbed Laplacian, we introduce the perturbation parameter
$\varepsilon = \frac{1}{\gamma}$ and we get the eigenvalue equation
\begin{equation}
	\left(\varepsilon L_{i j}+\Pi_{i j}^{\|}\right)\left(v_{j} + \varepsilon \delta v_{j}\right)=(1+\varepsilon \delta \lambda)\left(v_{i}+\varepsilon \delta v_{i}\right)
\end{equation}
Recalling that $\Pi_{i j}^{\|} v_{j} = \delta L v_{j} = v_{j} $ and $v_i=0$ for $i\neq k,h$, it follows
\begin{equation*}
	L_{i j}\left(v_{j}+\varepsilon \delta v_{j}\right)+\Pi_{i j}^{\|} \delta v_{j}=\delta \lambda v_{i}+(1+\varepsilon \delta \lambda) \delta v_{i}.
\end{equation*}
Projecting the above equation to the $N-1$ dimensional subspace orthogonal to the dipole state, we get
\begin{equation}\label{eq:perpendicular}
		\Pi^{\perp} L_{i j}\left(v_{j}+\varepsilon \delta v_{j}\right) =(1+\varepsilon \delta \lambda) \Pi^{\perp} \delta v_{i}.
\end{equation}
Whereas, projecting the same equation in the subspace parallel to the dipole state and setting $\Pi^{\|} \delta v_i=0$, we obtain
\begin{equation*}
	\delta \lambda v_{i} =  \Pi^{\|} L_{i j} v_{j}. 
\end{equation*}
Performing the scalar product with $v_i$, since $\Vert \vec v\Vert=1$ the eigenvalue change follows
\begin{equation*}
	\delta \lambda=v_{i} \Pi^{\|} L_{i j} v_{j}.
\end{equation*}
From equation (\ref{eq:perpendicular}) we have   
\begin{equation}\label{eq:pertv}
	\delta v_{i} =\frac{\Pi^{\perp} L_{i j}}{1+\varepsilon \delta \lambda}\left(v_{j} + \varepsilon \delta v_{j}\right)
\end{equation}
that allows to compute the perturbation $\delta v_{i}$ of the dipole eigenvector. Since the subspace with $\lambda = 1$ is one-dimensional, the perturbation to the dipole must be perpendicular to the dipole itself. Therefore, if $\varepsilon$ is small compared to the norm of $L$, the solution of eq. (\ref{eq:pertv}) exists and it satisfies
\begin{equation}\label{eq:residual coupling}
	\lim _{\varepsilon \rightarrow 0} \delta v_{i}=\Pi^{\perp} L_{i j} v_{j}
\end{equation}
The limit $\varepsilon \rightarrow 0$ means that the edge weight $w_{kh} \rightarrow \infty$, i.e. the two nodes $k$ and $h$ become infinitely coupled. Eq. (\ref{eq:residual coupling}) means that the perturbation of the dipole eigenvector, in the limit of an infinite weight $w_{kh}$ is given by the projection of the image of the dipole eigenvector through the Laplacian matrix e of $L_{i j} v_{j}$ on the orthogonal subspace: i.e. one realizes a constraint in the corresponding dynamics (\ref{eq:spring-mass eom}). 
Considering the other eigenvectors of $\delta L_{ij}$ (equation (\ref{eq:lambda 0}) with zero eigenvalue). We can choose $N-1$ independent arbitrary vectors belonging to that subspace, and we have
\begin{equation*}
	\left(\varepsilon L_{i j} + \Pi_{i j}^{\|}\right)\left(u_{j} + \varepsilon \delta u_{j} \right) = \varepsilon \mu\left(u_{i} + \varepsilon \delta u_{i} \right)
\end{equation*}
Since the subspace is $N-1$ degenerate, we take $\delta u_{j} = \alpha v_{j}$
\begin{equation*}
	\left(\varepsilon L_{i j}+\Pi_{i j}^{\|}\right)\left(u_{j}+\alpha \varepsilon v_{j} \right)=\varepsilon \mu\left(u_{i}+\alpha \varepsilon v_{i} \right)
\end{equation*}
and recalling that $\Pi_{i j}^{\|} u_j = 0$ and $\Pi_{i j}^{\|} v_j = v_i$ we obtain
\begin{equation}
	L_{i j} u_{j}+\alpha\left(v_{i} + \varepsilon L_{i j} v_{j} \right)=\mu u_{i}+\varepsilon \alpha \mu v_{i}.
\end{equation}
As before we project this equation on the subspace perpendicular and parallel to the dipole vector. In the first case we have the equation
\begin{equation*}
	\Pi^{\perp} L_{i j} u_{j} + \underbrace{\alpha \Pi^{\perp} v_{i}}_{=0} + \varepsilon \alpha \Pi^{\perp} L_{i j} v_{j} =  \underbrace{\mu \Pi^{\perp} u_{i}}_{=\mu u_i} + \underbrace{ \varepsilon \alpha \mu \Pi^{\perp} v_{i} }_{=0}
\end{equation*}
and, using $(\Pi^{\perp})^2 = \mathbb{I}$ and $\Pi^{\perp} u_i = u_i$ we get 
\begin{equation}\label{eq:perpe}
	\Pi^{\perp} L_{i j} \Pi^{\perp} u_{j} =\mu u_{i}-\varepsilon \alpha \Pi^{\perp} L_{i j} v_{j}.
\end{equation}
whereas, for the component parallel to the dipole we have
\begin{equation}\label{eq:paral}
	\alpha(1-\varepsilon \mu) =-v_{i}  L_{ij}\left(u_{j}+\varepsilon \alpha v_{j} \right)
\end{equation}
The equation system 
\begin{equation}
	\begin{cases}
		\Pi^{\perp} L_{i j} \Pi^{\perp} u_{j} &= \mu u_{i}-\varepsilon \alpha \Pi^{\perp} L_{i j} v_{j} \\
		\alpha(1-\varepsilon \mu) &= -v_{i}  L_{ij}\left(u_{j}+\varepsilon \alpha v_{j} \right)
	\end{cases}
\end{equation}
establishes the relation between the Laplacian properties of the reduced network and the original one. The first equation defines the Laplacian matrix of the reduced graph
\begin{equation}
	L_{ij} \rightarrow \Pi^{\perp} L_{i j} \Pi^{\perp}. 
\end{equation}
and the second equation determines the parameter $\alpha$. 
In the limit $\varepsilon\to 0$ we recover the new stationary solution, that is orthogonal to the dipole vector $\vec{v}$.\par\noindent
If during the limit $\varepsilon\to 0$ two eigenvalues of $\Pi^{\perp} L_{i j} \Pi^{\perp} $ are perturbed enough to collapse, the perturbation approach breaks down since a two dimensional invariant subspace is created, and any couple of independent vectors belonging to this subspace is an eigenvector base for the corresponding eigenvalue. As a consequence, in the limit $\varepsilon\to 0$, the spectrum of $\Pi^{\perp} L_{i j} \Pi^{\perp} $ may encounter bifurcations and the spectral properties of the reduced graph Laplacian $\Pi^{\perp} L_{i j} \Pi^{\perp} $ may be different from the ones of the original matrix $L$. In the dimensionality reduction procedure, one needs to keep the bifurcations under control. In general only the weights large enough can be put to infinity. \par\noindent
We remark that the projection operator $\Pi^{\perp}$ can be represented by a $N-1 \times N$ matrix whose rows are the eigenvectors (\ref{eq:lambda 0}). Therefore, it acts on a vector leaving unchanged its components different from $k$ and $h$, while it mixes the sub-spaces of nodes $k$ and $h$, summing them. This corresponds to the single entity generated by taking the weight $w_{kh}$ to infinity, i.e. a node that behaves effectively like the sum of the two original nodes $k$ and $h$. The analogy with a spring-mass system, can give further physical insights. In fact, the reduced Laplacian matrix $\Pi^{\perp} L_{i j} \Pi^{\perp} $, that is $N-1 \times N-1$ dimensional, represents a network similar to the original one, with the only difference that the nodes $k$ and $h$ are replaced by a single node, with double the mass of the original nodes (that were taken to be the same, since we used the un-normalized Laplacian matrix), and whose connections (springs) with the rest of the network are the sum of the original connections of nodes $k$ and $h$.
We recall that if the reduced network can be generated by \textit{external equitable partition}, it is also called \textit{quotient graph} and the spectral properties of the Laplacian matrix are preserved \cite{eep}.

%% file: bibliography.bib
@book{harary_social,
	title={Graph theory as a mathematical model in social science},
	author={Harary, Frank and Norman, Robert Z},
	number={2},
	year={1953},
	publisher={University of Michigan, Institute for Social Research Ann Arbor}
}

@inproceedings{wdn_spectral,
	author = {Di Nardo, Armando and Giudicianni, Carlo and Greco, Roberto and Herrera, Manuel and Santonastaso, Giovanni and Scala, Antonio},
	year = {2018},
	month = {07},
	pages = {},
	title = {Sensor Placement in Water Distribution Networks based on Spectral Algorithms},
	doi = {10.29007/whzr}
}

@book{dynamical_processes_on_complex_networks,
	 place={Cambridge}, title={Dynamical Processes on Complex Networks}, DOI={10.1017/CBO9780511791383}, publisher={Cambridge University Press}, author={Barrat, Alain and Barthélemy, Marc and Vespignani, Alessandro}, year={2008}
 }

@article{barabasi2016,
         author = {Gao, Jianxi and Barzel, Baruch and  Barabasi, Albert-László},
         year = {2016},
         pages = {307-312},
         volume = {530},
         title = {Universal resilience patterns in complex networks},
         journal = {Nature}
}

@article{daganzo2008,
title = {An analytical approximation for the macroscopic fundamental diagram of urban traffic},
journal = {Transportation Research Part B: Methodological},
volume = {42},
number = {9},
pages = {771-781},
year = {2008},
issn = {0191-2615},
doi = {https://doi.org/10.1016/j.trb.2008.06.008},
url = {https://www.sciencedirect.com/science/article/pii/S0191261508000799},
author = {Carlos F. Daganzo and Nikolas Geroliminis}
}

@article{kurata2010,
         author = {Inoue Kentaro and Li Weijiang and Kurata Hiroyuki},
         year = {2010},
         pages = {e12623},
         volume = {5(9)},
         title = {Diffusion Model Based Spectral Clustering for Protein-Protein Interaction Networks},
         journal = {PLoS ONE}
}

@article{moutsinas2020,
         author = {Moutsinas, Giannis and Guo, Weisi},
         year = {2020},
         pages = {3599},
         volume = {10},
         title = {Node-Level Resilience Loss in Dynamic Complex Networks},
         journal = {Sci Rep}
}

@article{yeon2023,
         author = {Yeon, H. and Eom, T. and Jang, K. and Yeo, J.},
         year = {2023},
         pages = {5154},
         volume = {13},
         title = {DTUMOS, digital twin for large-scale urban mobility operating system},
         journal = {Sci Rep}
}

@article{gonzalez2023,
         author = {Ambühl, Lukas and Menendez, Monica and González, Marta C.},
         year = {2023},
         pages = {26},
         volume = {6},
         title = {Understanding congestion propagation by combining percolation theory with the macroscopic fundamental diagram},
         journal = {Commun Phys}
}

@book{newman_networks_2018,
	title = {Networks},
	isbn = {0192527495},
	publisher = {Oxford University Press, 2018},
	author = {Newman, Mark},
	year = {2018}
}

@article{shuman_emerging_2013,
	title = {The emerging field of signal processing on graphs: {Extending} high-dimensional data analysis to networks and other irregular domains},
	volume = {30},
	number = {3},
	journal = {IEEE Signal Processing Magazine},
	author = {Shuman, D. I. and Narang, S. K. and Frossard, P. and Ortega, A. and Vandergheynst, P.},
	month = may,
	year = {2013},
	pages = {83--98},
}

@article{localization_laplacian,
	author = {Hata, Shigefumi and Nakao, Hiroya},
	year = {2017},
	month = {12},
	pages = {},
	title = {Erratum: Localization of Laplacian eigenvectors on random networks},
	volume = {7},
	journal = {Scientific Reports},
	doi = {10.1038/s41598-017-06298-6}
}

@article{fortunato_community_2010,
	title = {Community detection in graphs},
	volume = {486},
	number = {3-5},
	journal = {Physics Reports},
	author = {Fortunato, Santo},
	month = feb,
	year = {2010},
	pages = {75--174},
}

@article{G-N_algorithm,
	author = {M. Girvan  and M. E. J. Newman },
	title = {Community structure in social and biological networks},
	journal = {Proceedings of the National Academy of Sciences},
	volume = {99},
	number = {12},
	pages = {7821-7826},
	year = {2002},
}

@article{G-N_modularity,
	author = {Newman, Mark and Girvan, Michelle},
	year = {2004},
	month = {03},
	pages = {026113},
	title = {Finding and Evaluating Community Structure in Networks},
	volume = {69},
	journal = {Physical review. E, Statistical, nonlinear, and soft matter physics},
	doi = {10.1103/PhysRevE.69.026113}
}

@article{von_luxburg_tutorial_2007,
	title = {A tutorial on spectral clustering},
	volume = {17},
	number = {4},
	journal = {Statistics and Computing},
	author = {von Luxburg, Ulrike},
	month = dec,
	year = {2007},
	pages = {395--416},
}

@article{spectral_clustering_with_physical_intuition,
	title = {Spectral clustering with physical intuition on spring–mass dynamics},
	volume = {351},
	number = {6},
	journal = {Journal of the Franklin Institute},
	author = {Park, Jinho and Jeon, Moongu and Pedrycz, Witold},
	month = jun,
	year = {2014},
	pages = {3245--3268},
}

@article{eep,
	title = {Observability and coarse graining of consensus dynamics through the external equitable partition},
	author = {O'Clery, Neave and Yuan, Ye and Stan, Guy-Bart and Barahona, Mauricio},
	journal = {Phys. Rev. E},
	volume = {88},
	issue = {4},
	pages = {042805},
	numpages = {13},
	year = {2013},
	month = {Oct},
	publisher = {American Physical Society},
	doi = {10.1103/PhysRevE.88.042805},
	url = {https://link.aps.org/doi/10.1103/PhysRevE.88.042805}
}

@article{ornstein_U,
	author = {Vatiwutipong, Pat and Phewchean, Nattakorn},
	year = {2019},
	month = {07},
	pages = {},
	title = {Alternative way to derive the distribution of the multivariate Ornstein–Uhlenbeck process},
	volume = {2019},
	journal = {Advances in Difference Equations},
	doi = {10.1186/s13662-019-2214-1}
}

@misc{tao_1970,
 	 title={[PDF] topics in random matrix theory: Semantic scholar},
 	 url={https://www.semanticscholar.org/paper/Topics-in-Random-Matrix-Theory-Tao/611503ed9d36bad843374774c59ab335ebf7eac4}, journal={undefined}, author={Tao, T.}, year={1970}, month={Jan}
}

@article{hand_data_2008,
	title = {Data {Clustering}: {Theory}, {Algorithms}, and {Applications} by {Guojun} {Gan}, {Chaoqun} {Ma}, {Jianhong} {Wu}},
	volume = {76},
	number = {1},
	journal = {International Statistical Review},
	author = {Hand, David J.},
	month = apr,
	year = {2008},
	pages = {141--141},
}

@InProceedings{networkx,
	author =       {Aric A. Hagberg and Daniel A. Schult and Pieter J. Swart},
	title =        {Exploring Network Structure, Dynamics, and Function using NetworkX},
	booktitle =   {Proceedings of the 7th Python in Science Conference},
	pages =     {11 - 15},
	address = {Pasadena, CA USA},
	year =      {2008},
	editor =    {Ga\"el Varoquaux and Travis Vaught and Jarrod Millman},
}

@article{nonlocal_impact_of_link_failure,
	doi = {10.1088/1367-2630/ab13ba},
	url = {https://dx.doi.org/10.1088/1367-2630/ab13ba},
	year = {2019},
	month = {may},
	publisher = {IOP Publishing},
	volume = {21},
	number = {5},
	pages = {053009},
	author = {Julius Strake and Franz Kaiser and Farnaz Basiri and Henrik Ronellenfitsch and Dirk Witthaut},
	title = {Non-local impact of link failures in linear flow networks},
	journal = {New Journal of Physics},
}

@article{spatial_networks,
	title = {Spatial networks},
	journal = {Physics Reports},
	volume = {499},
	number = {1},
	pages = {1-101},
	year = {2011},
	issn = {0370-1573},
	doi = {https://doi.org/10.1016/j.physrep.2010.11.002},
	url = {https://www.sciencedirect.com/science/article/pii/S037015731000308X},
	author = {Marc Barthélemy},
}

@article{edge-based-elastic,
	title = {Edge-based formulation of elastic network models},
	author = {Hodges, Maxwell and Yaliraki, Sophia N. and Barahona, Mauricio},
	journal = {Phys. Rev. Research},
	volume = {1},
	issue = {3},
	pages = {033211},
	numpages = {9},
	year = {2019},
	month = {Dec},
	publisher = {American Physical Society},
	doi = {10.1103/PhysRevResearch.1.033211},
	url = {https://link.aps.org/doi/10.1103/PhysRevResearch.1.033211}
}

@article{prediction-of-allosteric-sites,
	author = {Amor, Benjamin and Schaub, Michael and Yaliraki, Sophia and Barahona, Mauricio},
	year = {2016},
	month = {08},
	pages = {12477},
	title = {Prediction of allosteric sites and mediating interactions through bond-to-bond propensities},
	volume = {7},
	journal = {Nature Communications},
	doi = {10.1038/ncomms12477}
}

@article{mason_graph_2007,
	title = {Graph theory and networks in {Biology}},
	volume = {1},
	issn = {1751-8849, 1751-8857},
	url = {https://digital-library.theiet.org/content/journals/10.1049/iet-syb_20060038},
	doi = {10.1049/iet-syb:20060038},
	number = {2},
	journal = {IET Systems Biology},
	author = {Mason, O. and Verwoerd, M.},
	month = {mar},
	year = {2007},
	pages = {89--119}
}

@article{cascade_failure_Mirzasoleiman,
  title = {Cascaded failures in weighted networks},
  author = {Mirzasoleiman, Baharan and Babaei, Mahmoudreza and Jalili, Mahdi and Safari, Mohammad Ali},
  journal = {Phys. Rev. E},
  volume = {84},
  issue = {4},
  pages = {046114},
  year = {2011},
  month = {Oct},
  publisher = {American Physical Society}
}

@article{cascade_failure_Chen,
  title = {Nonlinear model of cascade failure in weighted complex networks considering overloaded edges},
  author = {Chen, Chao-Yang and Zhao, Yang and Gao, Jianxi and Stanley, Harry Eugene},
  journal  = {Scientific Reports},
  volume   =  {10},
  number   =  {1},
  pages    = {13428},
  month    =  {aug},
  year     =  {2020}
}

@article{network_susceptibilities,
  title = {Network susceptibilities: Theory and applications},
  author = {Manik, Debsankha and Rohden, Martin and Ronellenfitsch, Henrik and Zhang, Xiaozhu and Hallerberg, Sarah and Witthaut, Dirk and Timme, Marc},
  journal = {Phys. Rev. E},
  volume = {95},
  issue = {1},
  pages = {012319},
  numpages = {13},
  year = {2017},
  month = {Jan},
  publisher = {American Physical Society},
  doi = {10.1103/PhysRevE.95.012319},
  url = {https://link.aps.org/doi/10.1103/PhysRevE.95.012319}
}
